\documentclass[alpha-refs]{wiley-article}

\usepackage{amsmath,amssymb,amsfonts}
\usepackage{booktabs}
\usepackage{pdflscape}
\usepackage{lineno}
\usepackage{color}
\usepackage{ulem}
\usepackage{siunitx}
\usepackage[colorlinks=true, linkcolor=blue, citecolor=blue, urlcolor=magenta]{hyperref}

\papertype{Original Article}

\title{Dynamic and thermodynamic contributions to future extreme-rainfall intensification: a case study for Belgium}

\author[1\authfn{1},2\authfn{2}]{Jozefien Schoofs}
\author[1\authfn{1},3\authfn{3}]{Kobe Vandelanotte}
\author[1\authfn{1}]{Hans Van de Vyver}
\author[3\authfn{3}]{Line Van Der Sichel}
\author[3\authfn{3}]{Matthias Vandersteene}
\author[2\authfn{2}]{Fien Serras}
\author[2\authfn{2}]{Nicole P. M. van Lipzig}
\author[1\authfn{1},4\authfn{4}]{Bert Van Schaeybroeck}

\corraddress{Bert Van Schaeybroeck, Department of Meteorological and Climatological Research, Royal Meteorological Institute, Ringlaan 3, 1180 Brussels, Belgium}
\corremail{bertvs@meteo.be}

\fundinginfo{Belgian Science Policy (BELSPO) under Contract B2/223/P1/CORDEXbeII.}

\affil[1]{Department of Meteorological and Climatological Research, Royal Meteorological Institute, Brussels, 1180, Belgium}

\affil[2]{Department of Earth and Environmental Sciences, KU Leuven, Leuven, 3000, Belgium}

\affil[3]{Department of Physics and Astronomy, Ghent University, Ghent, 9000, Belgium}

\affil[4]{Department of Geography, Ghent University, Ghent, 9000, Belgium}

\begin{document}

\maketitle
% no more than 300 words
\begin{abstract}
Extreme precipitation is projected to become more frequent and more intense due to climate change and associated thermodynamical effects, but the local response of atmospheric circulation under future climate scenarios remains uncertain due mainly to dynamical differences. In this study, we outline a methodology for a regional assessment of future extreme precipitation based on the Lamb Weather Type classification and to evaluate future changes in weather patterns. While anticyclonic days occur most frequently over Belgium, extreme rainfall is mostly associated with days of cyclonic, westerly and south-westerly weather patterns. GCMs from CMIP6 are first selected based on their reliability in representing local atmospheric circulation patterns during days with extreme rainfall days. It was found that for our case study over Belgium, the future (end-of-the-century SSP3-7.0) changes in intensity and likelihood of rainfall extremes can be primarily attributed to thermodynamic factors, with minimal contribution from changes in atmospheric dynamics. Both intensity and probability of extreme rainfall increase for all seasons. While extreme-rainfall probabilities mostly increase in fall and winter, the associated intensity changes are dominated by positive changes in spring and summer. Additionally, the weather patterns that are historically associated with extreme rainfall, disproportionally contribute to these changes, especially to thermodynamic changes. More specifically, robust changes arise from an increased extreme-rainfall occurrence probability in case of cyclonic, south-westerly and westerly circulation types.
\end{abstract}
%\linenumbers

%\begin{multicols}{2}

\section{Introduction}\label{sec_intro}
Extreme rainfall events have become more frequent in Western and Central Europe and are very likely to increase further as climate change continues~\citep{IPCC2023}. Its intensity is projected to increase throughout the world in the future warming climate, the rate of which is dependent on the scenario considered~\citep{Hansen2024}. As extreme precipitation events can lead to flooding, landslides, and other natural disasters, they will significantly impact different aspects of the environment, society, and economy~\citep{IPCCatlas,IPCCWG2}. In recent years, several catastrophic events took place over Europe where warming rates are higher than the global one. In mid July 2021, Germany, Belgium, Luxemburg, and the Netherlands were hit by 3 days of excessive rainfall leading to more than 200 fatalities and enormous infrastructural damage~\citep{Tradowsky2023}. In September 2024, the storm Boris caused record-breaking rainfall and devastating floods over a wide area over Central and Eastern Europe~\citep{Athanase2024}. \cite{Athanase2024} identified an additional 9\% of rainfall and an 18\% larger area affected by the storm due to human-induced warming.
A month later, in October 2024, another record-breaking downpour lead to flash floods over Sevilla (Spain), ensuing again over 200 fatalities~\citep{WWA2024}. In a rapid analysis, the World Weather Attribution working group estimated that this rainfall was approximately 12\% heavier and twice as likely compared to a preindustrial climate. Flood risks are generally exacerbated by a high degree of urbanization, as is present in Belgium, associated with an enhanced degree of surface runoff and reduced groundwater recharge and evapotranspiration~\citep{Mees2015}. In Belgium, for instance, urban expansion was shown to lead to a 20\% increase in runoff between 1976 and 2000, a trend which is further intensifying~\citep{Poelmans2010}.

It is well established that atmospheric circulation plays a determining role in driving precipitation variability over Europe~\citep{pattison2011relationship,planchon2009application,Brisson2011,whitford2024atmospheric}. Several methods have been proposed to categorize these atmospheric circulation types (CTs)~\citep{huth2008classifications} and several works have investigated the historical CT connection with climatic variables such as rainfall~\citep{Brisson2011}, extreme precipitation~\citep{tramblay2013non}, heatwaves~\citep{Serras2024}, floods~\citep{Pattison2012}, and weather extremes in general~\citep{Pfahl2014,faranda2023atmospheric}. 

CTs also allow one to better understand regional climate changes. While, as expressed by the Clausius‐Clapeyron equation, warmer air can hold more moisture and therefore extreme rainfall is expected to be more intense~\citep{neelin2022precipitation}, changes in the CTs occurrence frequencies may oppose or strengthen this trend regionally. \cite{Hansen2024} focus on the relation between CTs and extreme precipitation and find it is stable under climate change over Scandinavia. \cite{Herrera-Lormendez2023} examined shifts in future CTs under a high-emission scenario and their impact on average summer precipitation and droughts. Finally, similar to the study of~\cite{cattiaux2013towards} for cold extremes,~\cite{Otero2018} use CTs to disentangle future changes in temperature into changes related to CT-frequency changes (i.e. dynamical changes) and temperature changes within each CT (i.e. thermodynamical changes). As far as we are aware, however, the role of the dynamical and thermodynamical contributions to changes in extreme-rainfall frequency and intensity and their seasonal dependence over Central Europe have not been investigated.

In line with observations, Global Circulation Models (GCMs) consistently show the large-scale intensification of precipitation extremes in a warming climate. However, projecting these changes at regional and local scales remains challenging~\citep{Hansen2024} especially their interpretation in terms of physical mechanisms~\citep{OGorman2015,neelin2022precipitation}. Additionally, despite their consistent improvements of the lastest generation of CMIP6 GCMs~\citep{CMIP6_eyring} with respect to CMIP5, GCMs suffer from difficulties in capturing regional circulation patterns~\citep{Brands2022,vautard2023heat}. In order to account for this, some works select models based on their reliability in capturing the historical climatological features of CT (e.g.~\cite{mcsweeney2015selecting,Serras2024}).

In this study, we focus on the CTs specifically associated with extreme-rainfall days to enhance regional projections.
The objective here is to better understand future changes in extreme precipitation over Belgium. Therefore we develop a methodology for regional assessment of future extreme precipitation, focusing on the influence of atmospheric circulation patterns, and to evaluate how changes in atmospheric circulations and thermodynamic factors will impact the intensity and likelihood of extreme rainfall events.
We therefore (i) identify the relationship between atmospheric circulation patterns and daily precipitation extremes, (ii) evaluate the GCM performance based on their reliability to reproduce atmospheric patterns linked to extreme precipitation, and, (iii) decompose the intensity and likelihood changes of extreme rainfall into dynamic and thermodynamic contributions through a case study over Belgium. To assess the relationship between CTs and extreme precipitation, a commonly-used circulation-classification methodology, the Lamb Weather Type (LWT) classification \citep{Serras2024,Brisson2011,Otero2018}, is used.

\section{Materials and methods}\label{sec_materials&methods}

\subsection{Study area}\label{subsec_studyarea}
\begin{figure}[h]
\centering
\includegraphics[width=0.6\textwidth]{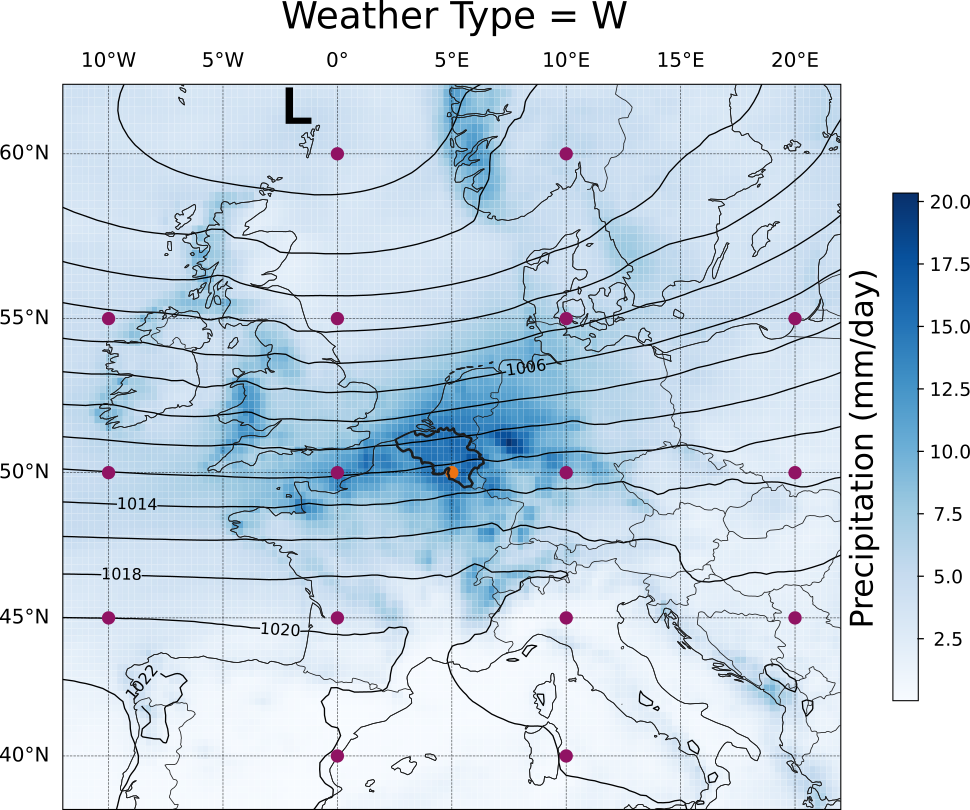}
\caption{Map of Western Europe with Belgium indicated with bold borders in the center. Mean sea level pressure from ERA5 at the 16 red-dotted points are used to determine the weather pattern over Belgium. Contours indicate the average pressure for the pattern with western flow (W) while the blue shading indicate the average ERA5 rainfall during extreme-rainfall days with western flow.}\label{fig_map}
\end{figure}

Belgium is situated in Western Europe (see Fig.~\ref{fig_map}) and features a temperate maritime climate, strongly influenced by the North Atlantic Ocean and characterized by relatively mild temperatures with little variation throughout the year~\citep{Brisson2011}. Precipitation patterns over Belgium are heavily affected by large-scale atmospheric circulation, including the North Atlantic Oscillation, especially in wintertime. Fronts from the west, which are associated with cyclonic events, are responsible for most of the precipitation.
Belgium is highly populated, urbanized, and industrialized, making it particularly vulnerable to extreme weather events such as heat waves, thunderstorms, and different types of flooding~\citep{Termonia2018, kmi_klimaatrapport}. Additionally, Belgium's varied orography has a notable effect on both average precipitation and extreme weather events~\citep{Wyard2017, Journee2015, Vandevyver2012, VandenBroucke2018} with annual average rainfall that varies from 600 mm/year near the coast to 1400 mm/year in the orographic regions~\citep{KMIatlas2024}.

%\end{multicols}

%\begin{multicols}{2}

\subsection{Data}

Observed precipitation data were used to investigate the relationship between the weather types and past extreme precipitation. Daily rain-gauge data were used for the period 1951-2023 at the station of Uccle (Belgium), managed by the Royal Meteorological Institute of Belgium.

To determine the weather types, we use ERA5, an atmospheric global reanalysis that embodies a detailed record of the global atmosphere, land surface, and ocean waves from 1940 onwards~\citep{ERA5_hersbach}. ERA5 is the fifth and latest generation of reanalysis produced by the European Centre for Medium-Range Weather Forecasts (ECMWF). Daily mean sea level pressure (MSLP) at a resolution of 0.25°$\,$x$\,$0.25° over Europe from 1951 to 2023 was used to establish the LWTs.
For the model evaluation, the period 1985-2014 is used as this coincides with the last 30 years of the CMIP6 historical simulations.

To evaluate potential changes in extreme precipitation due to climate change, a set of state-of-the-art coupled global climate models (GCMs) from the CMIP6 project is used. CMIP is a project of the World Climate Research Programme (WCRP)’s Working Group of Coupled Modelling~\citep{CMIP6_eyring}. An ensemble of 24 models with one member per model is considered since different ensemble members of GCMs are expected to feature similar atmospheric circulation patterns~\citep{Belleflamme2013}, hence giving equal weight to each model. Following~\cite{Serras2024}, different model families are used to obtain the widest possible range of model diversity. An overview of the selected CMIP6 models is given in Table~S1. We already excluded the CMIP6 models that were the least effective in reproducing the LWTs over Belgium during the summer-time period , as this is the most important period for extreme rainfall~\cite{Serras2024}.

For each model, we utilize the historical or reference period 1985-2014, and the future period 2069-2098 of the Shared Socioeconomic Pathways SSP3-7.0. This pathway represents a medium-to-high forcing scenario (i.e.~a radiative forcing level of $7\, \textnormal{W}/\textnormal{m}^2$ in 2100) under a regional rivalry scenario (i.e.~SSP3)~\citep{Guo2022} and it is chosen because of its expected higher signal-to-noise ratio. The 30-year periods were chosen in line with the guidelines of the WMO.

\subsection{Definition of extreme rainfall}\label{subsec_method_EP}

We define extreme rainfall using a well-established quantile-based method~\citep{vautard2016attribution} to fix a threshold. Under current climate conditions, a day with extreme precipitation is defined as a day with 20 mm/day or more of rainfall~\cite{vmm_extremeprecip}. Similarly, we identify extreme-rainfall days for CMIP6 model output based on the quantile of daily precipitation that corresponds to this 20 mm/day in the observations, i.e. the 98.87th quantile, within the historical period. A day with extreme precipitation in the CMIP6 model (both historical and future) is then defined as a day with rainfall that exceeds this model-dependent threshold. By using this ``quantile-mapping'' method, consistency in identifying days with extreme rainfall across both observed and modeled data is ensured. Finally, the extreme-rainfall intensity is the rainfall amount above the threshold on a day of extreme rainfall. Its average is therefore not an average over all days but only over those with extreme rainfall.

%\end{multicols}

\begin{table}[h!]
\caption{Overview of the two vorticity, eight directional, plus low flow (LF) Lamb Weather Types (LWTs), their acronym, and description, as well as their precipitation classification for Belgium. The wet and dry days are defined based on the occurrence probabilities shown in Fig.~S1.}\label{tab_wtexpl}%
\begin{tabular}{@{}l p{7cm} p{1.8cm} @{}}
\toprule
\\
Weather Type & Description & Wet/Dry \\
\midrule
\textit{Vorticity types} & & \\
Anticyclonic (A)    & Anticyclones centered over, near, or extending over Belgium & Dry \\
Cyclonic (C)        & Low pressure centered stagnating over, or frequently passing across Belgium & Wet \\
\midrule
\\
\textit{Directional types} & & \\
Easterly (E)        & Anticyclones north of Belgium, between Scandinavia and Iceland, and cyclones south of Belgium, in the Azores-Spain-Biscay region & Dry \\
Northerly (N)       & High pressure to the west and north-west of Belgium, and low pressure to the east and north-east of Belgium & Wet \\
North-easterly (NE) & Anticyclone over Scandinavia and cyclones south of Belgium & Dry \\
North-westerly (NW) & Anticyclone of the Azores displaced north to north-east towards Belgium, and depressions located north-east to Belgium & Wet \\
Southerly (S)       & High pressure covering central and northern Europe, North Atlantic depressions west of Belgium & Dry \\
South-easterly (SE) & Anticyclones over central or eastern Europe & Dry \\
South-westerly (SW) & Low pressure systems over the North Atlantic & Wet \\
Westerly (W)        & High pressure to the south and low pressure to the north of Belgium & Wet \\
\midrule
Low Flow (LF)       & Weak pressure gradients covering Europe & Dry \\
%\botrule
\end{tabular}
\end{table}

%\begin{multicols}{2}

\subsection{Lamb Weather Types}\label{subsec_method_LWT}

In this study, we use the objective circulation-type classification methodology introduced by~\cite{lamb1972}, and later adjusted by~\cite{Jenkinson1977}, known as the LWT classification or the ``Jenksinson–Collison Weather Type'' classification. For a detailed explanation of the original LWT classification algorithm we refer to~\cite{Serras2024}.
The circulation pattern at a given day is characterized by the positioning of high and low-pressure centers which control the orientation of the geostrophic airflow~\citep{Otero2018}. 
The LWT classification uses the MSLP on a 16-point grid with a footprint of 30° longitudes by 20° latitudes surrounding the center point at 50°N and 5°E (in Belgium) for which the LWT is defined.
From the MSLP of the 16 grid points, 6 indices are calculated that characterize the direction and vorticity of the geostrophic flow.  That is, the westerly and southerly flow components describe the area's primary zonal and meridional airflow, while total shear vorticity indicates atmospheric rotation. The classification of the LWT depends on the balance between the overall flow and the total vorticity.

In total, there are 27 circulation types including 8 directional types, 2 vorticity types, 16 hybrid types, and 1 unclassified type. Only the eight directional types (N, NE, E, SE, S, SW, W, NW), the two non-directional vorticity types (cyclonic C and anticyclonic A), and the unclassified weather type (Low Flow, LF) are included in this study following the approach of~\cite{demuzere2009analysis}. Table~\ref{tab_wtexpl} tabulates the synoptic situation over Belgium for each weather type. In this study, a distinction is made between wet and dry LWTs for Belgium (Table~\ref{tab_wtexpl}). Days with more than 1mm are classified as wet. A weather type, on the other hand, is considered wet (dry) if its occurrence probability is higher (lower) on a rainy day than on an average day (Fig.~S1). Note that the same classification of LWTs into wet and dry types is found if one uses days with extreme rainfall instead of wet days (see Fig.~S1). For illustrative purposes, we show in Fig.~\ref{fig_map} the average ERA5 MSLP and extreme-rainfall intensity during days of western flow.

Finally, note that not all CMIP6 GCMs have days of extreme precipitation for all weather types. For example, seven models did not project extreme precipitation for weather type SE, whereas for the cyclonic weather type C, all models did. Additionally, not all models project extreme precipitation to be present with weather types A and SE in the future period considered. Only 5 out of 19 models have days of extreme rainfall during weather type A, and two for SE. We tabulate all weather types without days of extreme rainfall per GCM in Table~S1.

\subsection{Circulation-based model evaluation and selection}\label{subsec_method_modelevaluation}

We evaluate the reproducibility of LWT probability during days of extreme rainfall for CMIP6 models as compared to ERA5 over the reference period (1985-2014). Based on this assessment, the worst-performing models are eliminated from further analysis. 

Similar to~\cite{Serras2024}, the model skill was quantified using the Perkins Skill Score (PSS). This metric quantifies the similarity between two probability density functions~\citep{Perkins2007} for which we consider the LWT probability distributions of ERA5 ($O$) and the CMIP6 models ($M$):
\begin{equation}
PSS = \sum_i \text{min} \left( P_{M,i}, P_{O,i} \right),\label{eq_PSS}
\end{equation}
where $i$ is the LWT, $P_{M,i}$ and $P_{O,i}$ the probability of LWT $i$ during days of extreme rainfall of the model and reference distribution, respectively.
If a model perfectly reproduces the reference probabilities, PSS is 1 while a PSS of 0 is obtained in case of a complete mismatch. Based on their PSS, models are ranked, and those with a PSS value below the mean minus one standard deviation, are excluded from further analysis, as in~\cite{Serras2024}.

\subsection{Decomposition into dynamic and thermodynamic contributions}

The goal here is to understand the future increase in extreme rainfall frequency and intensity over Belgium in terms of changes in ``dynamics'' and ``thermodynamics''. Assume a meteorological variable $G(d)$, that can be evaluated for each day $d$ (for instance, daily rainfall) and for which one is interested in its average change $\Delta \overline{G} = \overline{G}^f - \overline{G}^h$ between the future ($f$) and historical ($h$) period and the overline denotes an average over all days. This change can be written as a sum over LWTs $i$, with a proof given in Appendix B~\citep{Barry1973,vautard2016attribution,Souverijns2016,Otero2018}:
\begin{equation}
\Delta \overline{G} =\sum_i (\, \underbrace{ \overline{G}_{i}^f \Delta P_{i} }_\text{Dyn.}+ \underbrace{P_{i}^h \Delta \overline{G}_{i}}_\text{Thermodyn.}),
\label{eq_decomposition}
\end{equation}
where $\overline{G}_{i}$ denotes the average of $G(d)$ over days of weather type $i$, $P_{i}$ the probability of LWT $i$, and $\Delta$ indicates the difference between the future and historical value.
The first part of the equation represents the dynamic change induced by changes in the occurrence probability of the weather types. The second part, henceforth denoted as the thermodynamic contribution, quantifies the changes of $G(d)$, averaged within each weather type. Apart from changes in the large-scale thermodynamics, this term, however, may also encompass local or mesoscale feedbacks, or (despite its naming) even dynamic changes outside of the European domain that affect moisture advection towards Europe~\citep{Souverijns2016}.

\subsection{Changes in extreme-rainfall likelihood and intensity}
Assuming that $G(d)=1$ when there is extreme rainfall on day $d$ and $G(d)=0$ on other day, the period average $\overline{G}$ quantifies the likelihood of extreme rainfall within that period. The change in days of extreme rainfall, or, likelihood change is denoted as $\Delta P_{ext}$. The associated dynamic and thermodynamic changes are then associated with changes in LWT frequencies (over all day), and the changes in extreme-rainfall likelihood per LWT, respectively. 

The extreme-rainfall intensity is assumed to be the average excess rainfall on days of extreme rainfall. In other words, $G(d)$ is the rainfall amount minus the threshold and is only defined on days of extreme rainfall. Therefore period averages are restricted to days with extreme rainfall. The dynamic and thermodynamic changes in rainfall intensity are therefore 
caused by LWT frequency changes and within-LWT intensity changes, respectively, \textit{but both only during days of extreme rainfall}. The amount of extreme rainfall days may, however, differ between the historical and future periods but these differences are expressed by the extreme-rainfall likelihood changes.

%\end{multicols}

\begin{figure}[h!]
\centering
\includegraphics[width=1\textwidth]{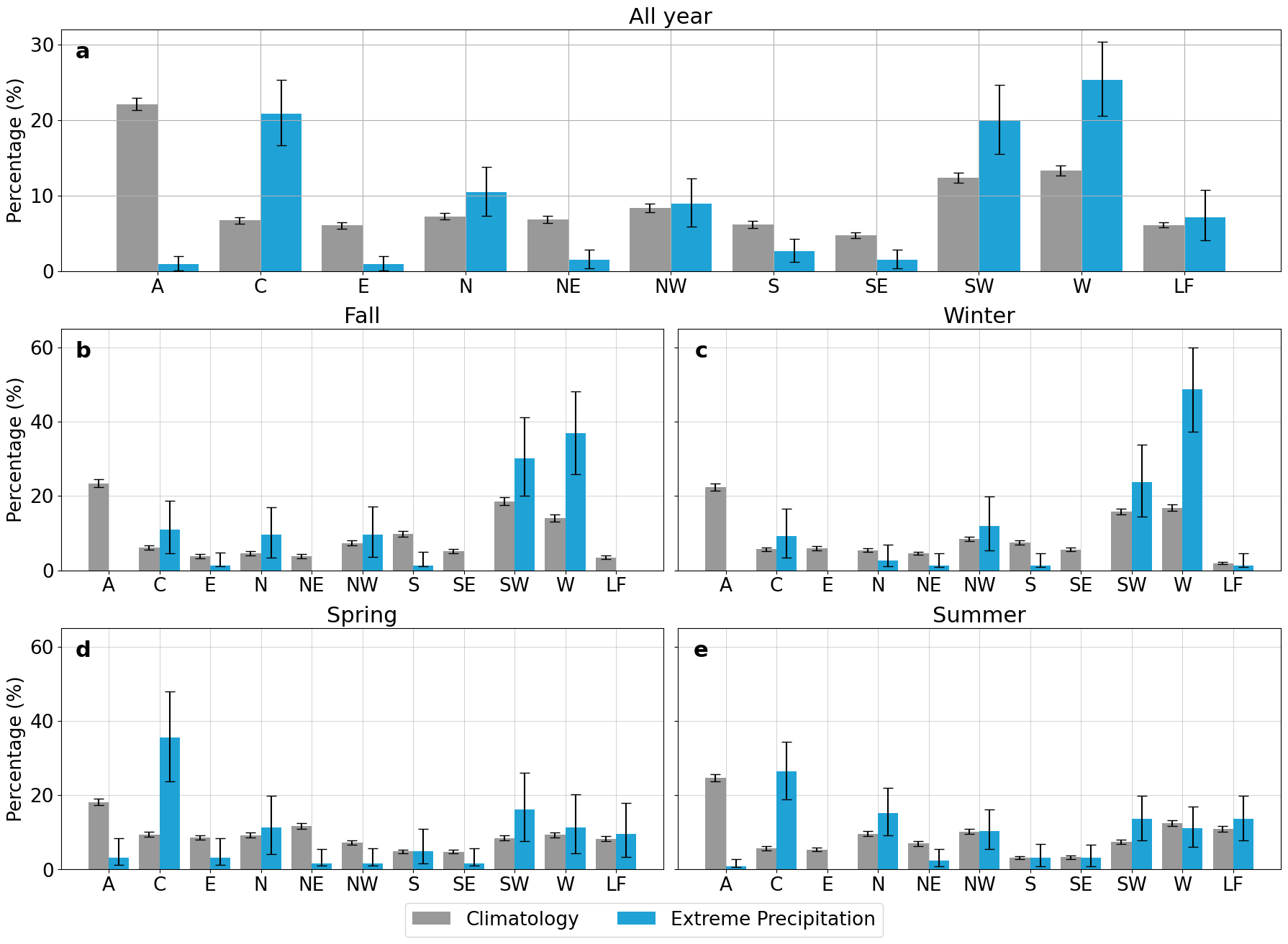}
\caption{Barplots representing the weather-type occurrence probability using ERA5 MSLP data and station precipitation at the grid box closest to Uccle (Belgium) during the period 1951-2023. The grey bars indicate the probability of weather types for all days (i.e.~the climatology) while the blue bars indicate the probability during days of extreme precipitation (which, in total, constitute about 1.13\% of all days). The top panel (a) presents the data for the whole year while the four seasons are presented in the panels (b-e). Error bars were created using $10^5$ bootstrap samples. The weather types are described in Table~\ref{tab_wtexpl}. }\label{fig_wtfrequency}
\end{figure}

%\begin{multicols}{2}

\section{Results}\label{sec_results}

\subsection{Observed weather-type distribution and link with extreme rainfall}\label{subsec_wtfreqanalysis}

Prior to analyzing future changes in extreme precipitation characteristics, we examine the current-day probability of the LWTs, for all days (i.e.~climatological probability), for wet days, and for days with extreme precipitation, based on the station observations at Uccle (Belgium). The weather-type probabilities in the historical period (1951-2023) are depicted in Fig.~\ref{fig_wtfrequency} and Fig.~S1. There are clear LWT probability differences among average days (grey bars), wet days (purple in Fig.~S1), and days with extreme precipitation (blue bars).

While anticyclonic (A) days occur most often over Belgium (i.e.~22.1\%), they rarely coincide with days of extreme rainfall (see Fig.~\ref{fig_wtfrequency}).
Likewise, the probability of LWTs E, NE, S, and SE is lower during days of extreme rainfall than their climatological probability, which is already low.
Therefore, the abovementioned dry LWTs (A, E, NE, S, and SE) coincide very rarely with days of extreme rainfall.

During days of extreme rainfall, the westerly (W) and south-westerly (SW) types occur most frequently, more specifically 25.3\% and 19.9\% (see Fig.~\ref{fig_wtfrequency}a). For W, this is 1.9 times higher than its climatological probability. 
Apart from W and SW, cyclonic (C) days are especially connected to extreme rainfall, since it features an occurrence probability of 20.8\%, while C has a low climatological probability of 6.7\%.
Furthermore, for Northern (N) flow, the occurrence probability during days of extreme rainfall exceeds the climatological probability while both probabilities more or less coincide for NW. Lastly, the probability of the low-flow (LF) LWT is similar between the two categories (5.7\% and 4.0\%).

Within the considered period (1951-2023), 3051 days of extreme rainfall were found, out of which 40\% (1250 cases) occurred during summer. Fall, winter, and spring, on the other hand, each account for 20\% of extreme-rainfall days, respectively. Figures~\ref{fig_wtfrequency}b-e show the variations in the LWT occurrence probability per season. Qualitatively, the extreme-rainfall probability in fall strongly resembles those in winter and those in spring resemble those in summer. Climatologically, anticyclonic days are the most prevalent in every season, reaching more than 20\% in each season (except in spring with 18.2\%). During days of extreme rainfall, on the other hand, LWT W is most prevalent in fall and winter (37.0\% and 48.7\%, respectively), and C in spring and summer (35.5\% and 26.4\%). 

The cyclonic weather type was substantially less associated with extreme precipitation in fall and winter as compared to spring and summer. In spring and summer C occured most frequently during days of extreme rainfall, i.e.~26.4\% in summer which is 4.6 times its climatological occurrence frequency. 

The 95\%-confidence intervals (CIs) of the LWT probabilities in Fig.~\ref{fig_wtfrequency} were calculated using $10^5$ resamples, obtained with a bootstrap procedure with replacement. As expected, due to the reduced sample sizes for days with extreme rainfall, the associated error bars are generally larger than those for the climatological probability.

\subsection{Model evaluation and selection}\label{subsec_modelevaluation}

In order to subselect models from the CMIP6 ensemble, we assess how the GCMs reproduce the circulation patterns over Belgium during days with extreme rainfall for the reference period (1985-2014) based on the PSS~\citep{Serras2024}. 
We evaluate the PSS for various variables, including the climatological LWT probability ($P_{clim}$) and the LWT probability during extreme-rainfall days ($P_{extr}$). The analysis revealed a low correlation (0.37, see Fig.~S2) between the PSS of the climatological LWT probability ($P_{clim}$) and that of the LWT probability during extreme-rainfall days ($P_{extr}$). These results suggest that a model's ability to accurately represent the climatological weather types is not necessarily an indicator of its skill in representing the LWT distribution during extreme precipitation events.

Figure~\ref{fig_modelevaluation} shows the LWT occurrence probabilities per model during days of extreme rainfall.
Overall, there is a reasonable agreement between the model and the ERA5 probabilities and the agreement is quantified using the PSS (see Eq.~\eqref{eq_PSS}) which is indicated in each panel. The PSS values range from 0.668 to 0.881 among the models and the panels in Fig.~\ref{fig_modelevaluation} are ranked from high to low PSS. 
The models therefore vary significantly in their accuracy of reproducing LWT probability during the reference period.
As aforementioned, the PSS threshold for model elimination is defined as the mean minus the standard deviation of the PSS which here is 0.697. Five out of 24 models (i.e. CanESM5, CESM2-WACCM, CESM2, IPSL-CM5A2-INCA, and TaiESM1) fall (slightly) below this performance threshold and are excluded from further analysis and are indicated in red in Fig.~\ref{fig_modelevaluation}. Three out of five lowest-performing models (i.e. CESM2-WACCM, CESM2, TaiESM1) originate from one model family, i.e.~CAM. Note, however, that the CanESM5 model was previously identified as the best model to represent the average circulation probability distribution over Belgium~\citep{Serras2024}.

TaiESM1, CESM2, and CESM2-WACCM strongly overestimate the occurrence frequency of the anticyclonic (A) weather type and underestimate the cyclonic (C) weather type during extreme precipitation. For CanESM5 and IPSL-CM5A2-INCA, the largest discrepancies are found for weather type W.
Caution has to be made that even the best-performing models are far from perfect. Here, the best-performing model GFDL-ESM4 has a PSS score of only 0.881. Additionally, MPI-ESM1-2-HR has a PSS score of 0.815 and rank 6. This model strongly overestimates LWTs SW and W, and underestimates LWTs C and N during extreme precipitation.
To conclude, the following climate-change analysis for Belgium will use only the 19 selected GCMs as the eliminated models may compromise reliability and accuracy.

%\end{multicols}

\begin{landscape}
\begin{figure}[h!]
\centering
\includegraphics[width=1.35\textwidth]{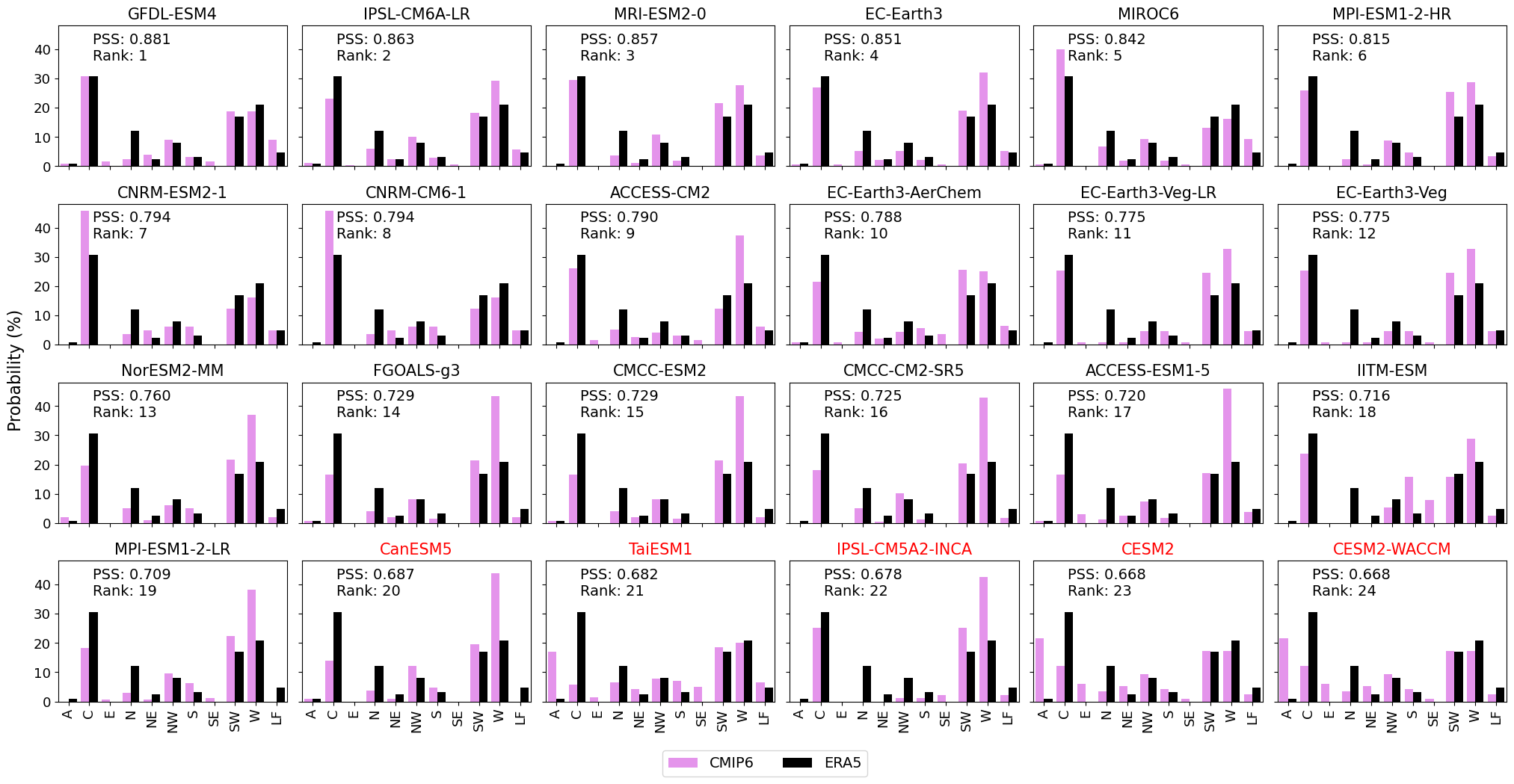}
\caption{Barplots representing the weather-type occurrence probability during days of extreme precipitation in Uccle (Belgium) for the different CMIP6 models (purple bars), compared against the reference data (ERA5, black bars) for the historical period 1985-2014. The Perkins Skill Score and its rank are indicated per model. Models with a PSS score below  0.697 are removed from the dataset for further analysis and are indicated in orange.}\label{fig_modelevaluation}
\end{figure}
\end{landscape}
\restoregeometry

\begin{figure}[h!]
\centering
\includegraphics[width=1\textwidth]{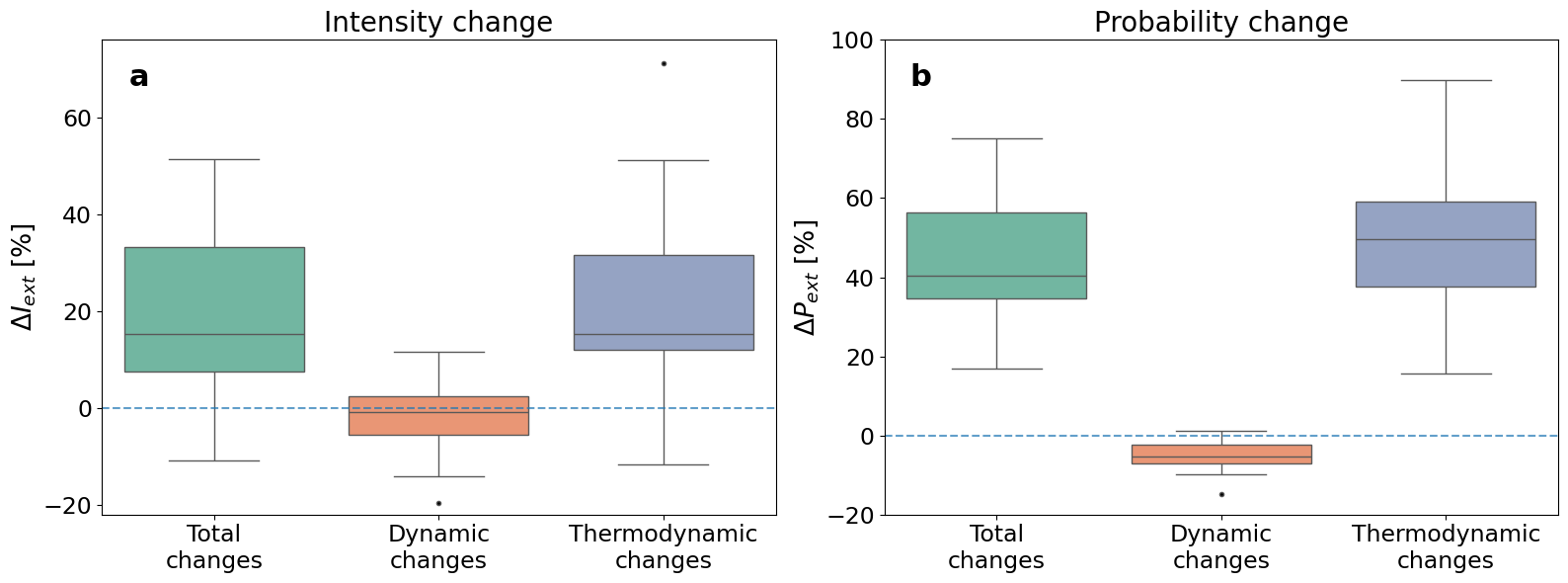}
\caption{Relative change in intensity (a) and occurrence probability (b) for days with extreme precipitation for 2069-2098 (following SSP3-7.0) with respect to 1985-2014, according to the 19 selected CMIP6 models. The green boxplot represents the total change in intensity, while the orange and blue indicate the dynamic and thermodynamic contributions, respectively. The median, interquartile range, and outliers of these changes over the CMIP6 models are represented by the solid black line, the box, and black dots, respectively.}\label{fig_decompositiontotal_relative}
\end{figure}

%\begin{multicols}{2}

\subsection{Total intensity and likelihood changes}\label{subsec_totalchange}

Figure~\ref{fig_decompositiontotal_relative}a shows the relative change in extreme-rainfall intensity (green boxplot) in the future period (2069-2098) compared to the reference period (1985-2014) for the SSP3-7.0 scenario. The median of the relative change is 15.4\% (green bar), which corresponds to 0.59 mm/day in absolute change (see Fig.~S3). There is a big spread among the different models in the magnitude of change with two out of 19 models showing an intensity decrease, more specifically, -10.8\% for NorESM2-MM and -1.6\% for IPSL-CM6A-LR, corresponding to reductions in intensity of 0.49mm/day and 0.14mm/day, respectively.
The CNRM-CM6-1 model features the highest relative change value of 51.4\%, an intensity increase of 2.28mm/day.

As seen in Eq.~\eqref{eq_decomposition}, the total changes can be split into the contributions of the dynamics and thermodynamics, indicated with orange and blue boxes in Fig.~\ref{fig_decompositiontotal_relative}a, respectively. By far the strongest contribution is attributable to the thermodynamics i.e.~changes in the
properties of weather types. The thermodynamic changes are 15.4\% (0.79mm/day) and this is a robust result as 95\% of the models
indicate a positive intensity change due to thermodynamics. The dynamic contribution cannot be considered as significantly different from zero.

Figure~\ref{fig_decompositiontotal_relative}b shows the total relative change (green) in occurrence probability of days with extreme rainfall and their decomposition into the dynamic change (orange) and thermodynamic changes (blue).
The median relative likelihood change is 40.3\% and all models feature positive changes, ranging from 16.9\% for CNRM-ESM2-1 to 75.0\% for EC-Earth3-Veg. In other words, while, by definition, the historical occurrence probability of days with extreme precipitation is equal to 1.1\% for all models (by definition, see Sect.~\ref{subsec_method_EP}), by the end-of-the-century following SSP3-7.0, this likelihood is expected to increase to 1.6\% on average.

As was the case for the intensity changes, the thermodynamic contribution is responsible for the largest positive changes in extreme-rainfall probability while the (absolute) dynamic changes are comparatively very small. However, as opposed to the dynamic contribution of the intensity changes, the dynamic contribution to the extreme-rainfall likelihood changes are small but robustly negative with 90\% of the models agreeing on the sign. More specifically, the dynamic contribution to the likelihood of days with extreme precipitation would decline by 0.06\%. The thermodynamic probability changes, on the other hand, are positive for all models, indicating a likelihood increase of days with extreme precipitation with 56\% (median value) due to within-type changes. As a result, thermodynamic changes have the largest contribution to the total changes in the likelihood.

%\end{multicols}

\begin{figure}[ht!]
\centering
\includegraphics[width=1\textwidth]{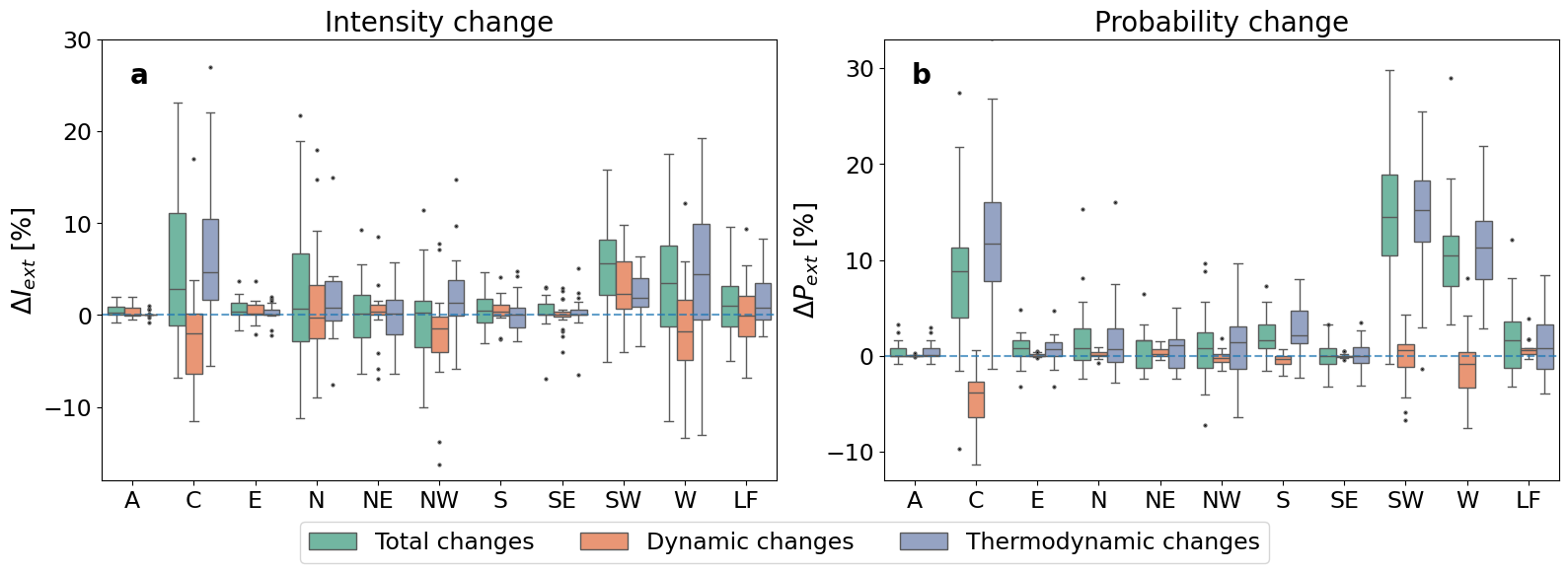}
\caption{Relative change in intensity (a) and occurrence probability (b) of extreme-precipitation days per weather type for 2069-2098 (following SSP3-7.0) with respect to 1985-2014, with bar plots showing the distributions of 19 CMIP6 models. The green boxplot represents the total relative change in intensity, and the orange and blue indicate the dynamic and thermodynamic contributions, respectively. The median, interquartile range, and outliers of these changes over the CMIP6 models are represented by the solid black line, the box, and black dots, respectively. Note that the \textit{relative} changes are relative with respect to the total intensity and total probability (not those per LWT). Furthermore, the y-axis of (a) is clipped between -18\% and 30\% masking the following outliers: For C, CNRM-CM6-1 has a dynamic contribution of -42.9\%, and a thermodynamic contribution of 55.8\%. The y-axis of (b) is clipped between -13\% and 33\% also masking outliers: IPSL-CM6A-LR has a total change of 45.2\%, of which a thermodynamic contribution of 37.0\%. CMCC-ESM2 has a thermodynamic contribution of 36.6\%.}\label{fig_decompositionwt_relative}
\end{figure}

%\begin{multicols}{2}

\subsection{Weather-type contributions under climate change}

\subsubsection{Intensity changes per LWT}
The contributions per LWT to the total relative changes in extreme-rainfall intensity are plotted in Fig.~\ref{fig_decompositionwt_relative}a in the green boxplots  (the absolute changes are shown in Fig.~S4a). All weather types show an increase in median intensity.

The largest total extreme-rainfall intensity increase is found for SW (5.6\% or 0.246mm/day) for which approximately 84\% of the models show a positive change in intensity. The projected intensity increases for weather types C, N, and W are comparatively as large but show larger uncertainties. The changes for the low-flow (LF) type have a near-zero median but are positively skewed. The smallest increases are associated with the weather types SE, NE, followed by NW, A, and E. For Belgium, the weather types A, E, NE, and SE are associated with dry weather. To elaborate, in the observed historical period (1985-2014) these weather types corresponded to 89.3\% (A), 93.7\% (E), and 91.5\% (SE) of dry days, as well as (almost) no extreme precipitation in the reference and future period. 

The dynamic and thermodynamic contributions to the intensity changes per weather type are also shown in Fig.~\ref{fig_decompositionwt_relative}a.
The wet weather types (i.e.~C, N, W, SW, see Fig~\ref{fig_wtfrequency}) occur most often for days with extreme precipitation, and show the largest range of dynamic and thermodynamic changes.
Their dynamic contributions decrease the extreme-rainfall intensity with about -1.98\%  (C), -0.28\% (N), and -1.74\%  (W). These changes are relatively reliable as 26\% of the models project negative changes for C, 37\% for W, and 47\% for N. The thermodynamic changes of these types are positive, with high model agreement for C and W (79\% and 74\%, respectively) but only 63\% for N. Given that the thermodynamic changes exceed the dynamic changes, the total changes for C, W and N are positive.
The south-westerly (SW) weather type has the largest total changes with positive dynamic and thermodynamic changes, featuring the highest dynamic changes of all LWTs (i.e.~2.27\% or 0.11mm/day), and a thermodynamic contribution of 1.89\% (0.11mm/day). It should be noted that for SW the uncertainty on the dynamic changes exceeds the one on the thermodynamic change.
For LWT NW, the dynamic intensity is negative and about -1.44\% or -0.066mm/day. These results are rather robust since 75\% of the models indicate this sign of change. However, due to large within-type changes (i.e.~thermodynamic changes) of 1.35\%, the total changes are still mostly positive (0.25\%).

The low-flow type, which featured equal occurrence probability during all days and days with extreme rainfall, shows only a modest positive thermodynamic contribution.
With respect to the changes of the aforementioned LWTs, the dry weather types A, S, and SE are not associated with substantial advection of (extreme) precipitation, not in the historical period, nor in the considered future period.

\subsubsection{Probability changes per LWT}
The total changes in the extreme-rainfall occurrence probability changes per weather type are presented as green boxplots in Fig.~\ref{fig_decompositionwt_relative}b (the absolute changes are shown in Fig.~S4b). The wet weather types C, SW, and W show positive increases (of magnitude 8.9\%, 10.5\%, and 14.5\%, respectively) that are much larger than those of the other LWTs. Additionally, the percentage of models indicating positive likelihood changes is also high, i.e.~89.0\% (C), 94.7\% (SW), and 100\% (W).
The remaining weather types show much smaller total changes, but all feature positive medians values. Their sign of change is nevertheless not uniformly positive among all models, e.g.~37\% of models indicate a reduction in occurrence probability in the considered future period for the weather types NW and LF.

The most prevalent weather types C, SW, and W during days of extreme precipitation (See Fig.~\ref{fig_wtfrequency}) also show the largest thermodynamic changes (red bars in Fig.~S4b), i.e.~11.7\% (C), 15.2\% (SW), and 11.3\% (W), with a strong agreement in sign among models, i.e.~95\% for C and SW, corresponding to 18 out of 19 models (i.e.~not for CNRM-CM6-1 for C, and GFDL-ESM4 for SW), and all 19 models for W. Dynamic changes of C and W, on the other hand, indicate a decrease in occurrence probability due to weather-type frequency changes, with a reduction of 3.9\% and 0.8\%, respectively. For SW, the change is slightly positive, with 68\% of models indicating a positive change.

Dry weather types like A, E, and SE, which historically are not related to extreme precipitation, show minimal dynamic changes, with small total changes primarily due to thermodynamics. Other dry types such as NE, S, and LF bring slightly more extreme precipitation, but changes in frequency and within-type intensity are minor. The wet types, NW and N show a wider range of changes, indicating greater model disagreement on the magnitude of change.

%\end{multicols}

\begin{figure}[h]
\centering
\includegraphics[width=\textwidth]{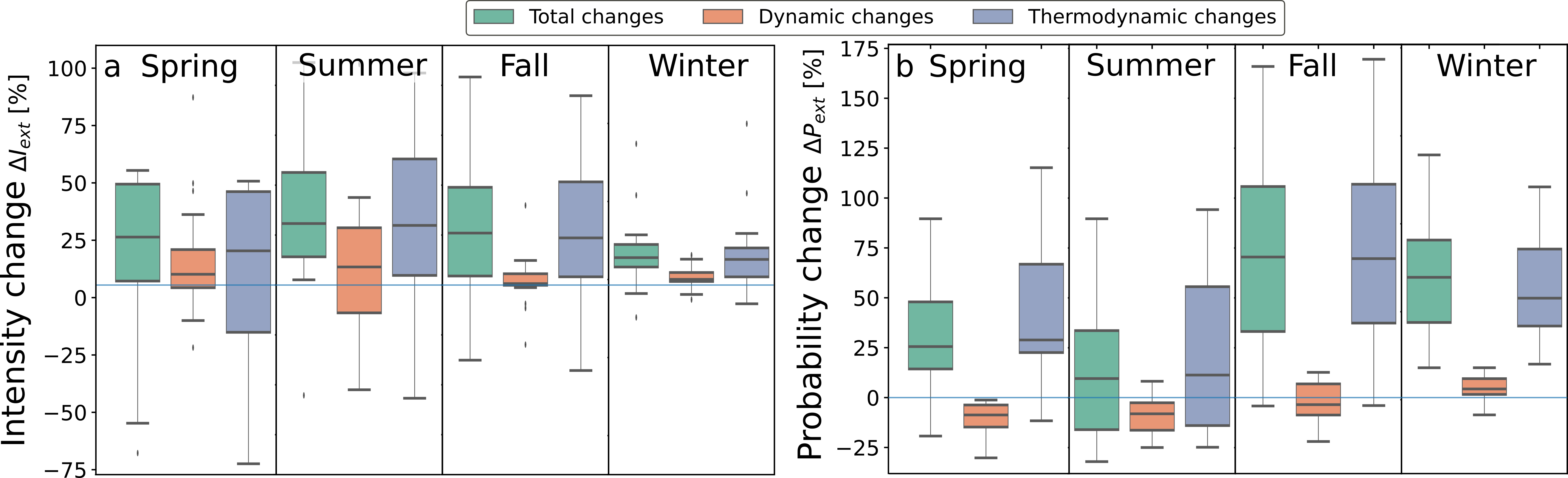}
\caption{Relative change in intensity per season for (a) and occurrence probability (b) for days with extreme precipitation for 2069-2098 (following SSP3-7.0) with respect to 1985-2014, according to the 19 selected CMIP6 models in Uccle, Belgium. The green boxplot represents the total change in intensity, while the orange and blue indicate the dynamic and thermodynamic contributions, respectively. The median, interquartile range, and outliers of these changes over the CMIP6 models are represented by the solid black line, the box, and black dots, respectively.}\label{fig_total_relative_decomposition_seas}
\end{figure}

%\begin{multicols}{2}

\subsection{Changes per season}
Figure~\ref{fig_total_relative_decomposition_seas}a shows the total, dynamic and thermodynamic changes for the extreme-rainfall intensity per season. During spring and summer, the total changes (green bars) are slightly higher than in fall and winter, however, with markedly high dynamical changes. The intensity changes are the smallest in winter. The seasonal contrasts are even more conspicuous for the changes in extreme-rainfall occurrence probability (Fig.~\ref{fig_total_relative_decomposition_seas}b) for which the biggest changes occur now in fall and winter and summer even features insignificant total changes. The dynamic changes are small as compared to the thermodynamic changes for all seasons and are significantly negative for spring and summer and significantly positive in winter.

Note that the uncertainties associated with the total changes per season indicated in Fig.~\ref{fig_total_relative_decomposition_seas} are frequently smaller than the ones of the thermodynamic contributions. This is indicative of compensation effects or negative correlations between the thermodynamic and dynamic contributions. These could arise due to future changes in the weather-type distribution that are decoupled from changes in extreme rainfall and may suggest that impact-tailored weather-type clustering methods (e.g.~\cite{Serras2024}) may be more suited to investigate these seasonal changes in extreme rainfall.

\begin{landscape}

\begin{figure}[h]
\centering
\includegraphics[width=1.4\textwidth]{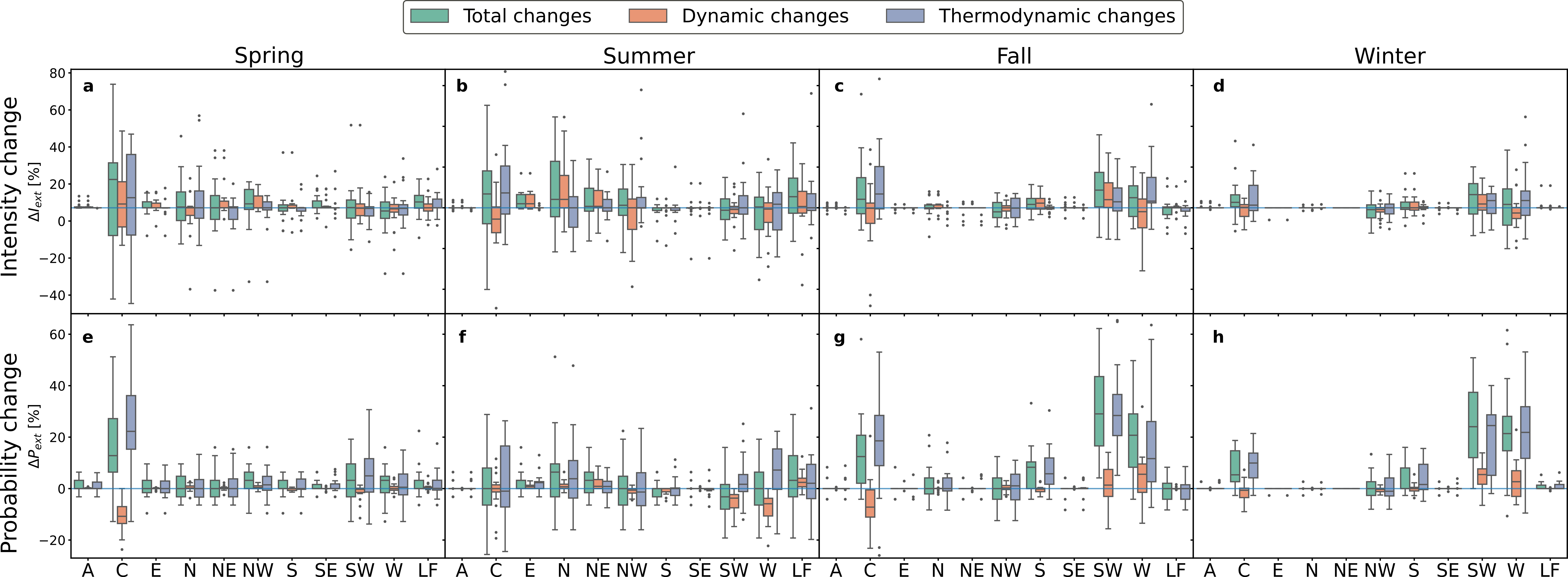}
\caption{Relative seasonal change in intensity (a-d) and occurrence probability (e-h) of extreme-precipitation days per weather type for 2069-2098 (following SSP3-7.0) with respect to 1985-2014 for Uccle (Belgium), with bar plots showing the distributions of 19 CMIP6 models. The green boxplot represents the total relative change in intensity, and the orange and blue indicate the dynamic and thermodynamic contributions, respectively. The median, interquartile range, and outliers of these changes over the CMIP6 models are represented by the solid black line, the box, and black dots, respectively. Note that the \textit{relative} changes are relative wrt to the total intensity and total probability (not those per LWT).}\label{fig_wt_relative_decomposition_seas}
\end{figure}

\end{landscape}

The contributions per weather type to the changes per season are shown in
Fig.~\ref{fig_wt_relative_decomposition_seas} where, again, the largest contributions arise from a few wet weather types (C, N, SW, W). While most changes are statistically insignificant for the intensity changes (Fig.~\ref{fig_wt_relative_decomposition_seas}a-d), some are for the frequency changes (Fig.~\ref{fig_wt_relative_decomposition_seas}e-h). More specifically, the large increases in extreme-rainfall probability in fall and winter are associated with both dynamic and thermodynamic increases for SW and W. The cyclonic weather type (C) also contributes significantly-positive total changes in fall and winter but also in spring. However, these positive total changes consistently involve large positive thermodynamic and small but significantly-negative dynamic contributions due to reductions in the cyclonic occurrence probability. 
(Fig.~\ref{fig_wt_relative_decomposition_seas}e,g,h).

%\end{multicols}

\restoregeometry

%\begin{multicols}{2}

\section{Discussion}\label{sec_discussion}

\subsection{Analysis of observed weather-type probability}\label{subsec_disc_observations}

We identified wet LWTs for Belgium including cyclonic (C), and the directional types westerly (W), and south-westerly (SW). In accordance with~\cite{Brisson2011} in which the analysis was performed using wet-day likelihood, the following weather types were defined as wet: northerly (N), and north-westerly (NW). These weather types are more prevalent during extreme-rainfall days compared to their climatological likelihoods in the considered reference period (1951-2023).
This finding broadly supports the work of other studies in this area (e.g.~\cite{Hellstrom2005} for Scandinavia and~\cite{Herrera-Lormendez2023} for Western Europe).
Weather types W, SW, and NW are often associated with precipitation as they transport warm maritime air from the North Atlantic~\citep{Herrera-Lormendez2023, Trigo2000, Brisson2011}. The western LWT is (W) most prevalent in fall and winter extreme precipitation, whereas it does not contribute much in spring and summer. In fall and winter, the North Sea and North Atlantic Ocean are relatively warmer than land due to the large relative heat capacity of water. As a result, maritime air cools down over land and a high degree of condensation may lead to (extreme) precipitation.
Furthermore, blocking anticyclonic events in regions to the northeast, northwest, and sometimes to the south of Belgium generally increases the likelihood of extreme rainfall, as described by~\cite{Lenggenhager2019}. These can correspond to the weather types NW, N, and W. 
\cite{Brisson2011} has shown that winter days are often wetter than days in other seasons but have a low average rainfall intensity over Belgium. In summer, in contrast, the rainfall intensity is higher. Therefore, in winter, long periods of persistent rainfall occur more often, whereas, in summer, extreme precipitation in shorter periods is more likely to be catastrophic.

Furthermore, the work in~\cite{Herrera-Lormendez2023} describes synoptic atmospheric patterns within the summer climatology. Over Europe, anticyclonic circulation (A) was found to be the most common, occurring on 20\% of days, supporting our findings of a 22\% year-round likelihood over Belgium.
In addition to A, also weather types E, NE, SE, and S which are connected to continental geostrophic flow, indicate dry conditions in Western Europe according to~\cite{Herrera-Lormendez2023} and in our results. This work also indicates that the low-flow conditions have no dominating impact on precipitation across most of Europe, due to a lack of moisture advection, which is in line with our findings.

Prior studies (e.g.~\cite{Schiemann2010}) have noted that circulation types more effectively capture mesoscale precipitation variability in mid-latitudes during winter as compared to summer. This might lead to a lower likelihood of wet LWTs for days with extreme rainfall during summer~\citep{Brisson2011}, consistent with our results.
In contrast, during winter extreme rainfall, the smallest variability within LWTs is found, with a dominance of weather type W.

\subsection{Model evaluation and selection}\label{subsec_disc_modelevaluation}

The study by~\cite{Brands2022} indicates that models in CMIP6 generally outperform their CMIP5 counterparts regarding the recurring regional atmospheric circulation patterns in global models.
In the study by~\cite{Li2021}, CMIP6 models are evaluated for their ability to capture daily rainfall extremes, defined as the upper tail quantiles of annual precipitation maxima, over various return periods (2 to 50 years) and various landmasses. They found reasonable accuracy in comparison to reanalysis data~\citep{Li2021}. However, these models still have some limitations.
\cite{Shepherd2014} identifies that uncertainties in atmospheric pattern projections stem mainly from natural internal variability, systematic model errors, and external influences such as volcanic and solar variability. The difficulty in accurately modeling small-scale processes affecting broader circulation patterns further contributes to these projection uncertainties~\citep{Shepherd2014}.

\cite{Brands2022} found a strong correlation between model performance in producing regional atmospheric circulation patterns in global models and horizontal resolution, with higher-resolution models generating better results. Moreover, they found different levels of performance accuracy across different regions. \cite{Brands2022} also suggest that the CMCC and NorESM2 model results are indicative for the CESM2 model results (for which the data was not available) as all derive from the NCAR model. This was, however, not confirmed by our analysis as, unlike the CMCC and NorESM2 models, CESM2 and CESM2-WACCM were discarded from the analysis and with distinct LWT biases during days of extreme rainfall (see Fig.~\ref{fig_modelevaluation}). This may be due to differences in convective parametrization schemes that are important for extreme rainfall. Additionally, contrary to our results for days of extreme rainfall,~\cite{Brands2022} found TaiESM1 to perform well over Europe. 

  Our work extends the methodology of ~\cite{Serras2024} to extreme rainfall. The threshold metric used is the Perkins Skill Score (PSS) which evaluates the distance of the LWT probability distribution during extreme-rainfall days between the model and the reference ERA5 (Fig.~S1). Apart from the climatological LWT probability ($P_{clim}$) and the LWT probability during days of extreme rainfall ($P_{extr}$), we also considered the PSS metric for the conditional probability of extreme rainfall given the occurrence of LWT $i$: ($P_{cond} \propto P_{extr}/P_{clim}$). The correlation between the PSS of the LWT probability fraction ($P_{cond}$) and the one of $P_{extr}$ is quite high ($0.82$, see Fig.~S6), while the one between $P_{cond}$ and $P_{clim}$ is low ($0.2$, see Fig.~S6). This indicates the dominant influence of the PSS during extreme-precipitation days on the PSS of $P_{cond}$. Therefore, we consider it justified to use the PSS score for days with extreme precipitation ($P_{extr}$) to distinguish models.

The GCM selection methodology in~\cite{Serras2024} limits the risk of choosing models with inadequate representation of the synoptic-scale dynamics. Their model assessment focused on the climatology for June, July, and August, and there were substantial differences in model rankings their and our analysis. 
This highlights the importance of evaluating models to determine their suitability for the intended application and is in line with the low correlation between the PSS of the climatological LWT distribution and the one during days of extreme rainfall.

\subsection{Intensity and likelihood changes and their decomposition}\label{subsec_disc_decomposition}

A strong agreement in the sign of change for extreme precipitation intensity has been reported in the literature at a global scale~\citep{IPCC2023}, with increases in intensity largely attributable to anthropogenic influence~\citep{Tabarietal2020,Tradowsky2023}.
CMIP6 climate projections, for example, foresee an increase in the number of very heavy rain days globally, with higher significance under higher SSP scenarios~\citep{Feng2023}.
While extreme precipitation will become more frequent, light-to-moderate precipitation may decline due to the drying and stabilizing effects of intense rainfall~\citep{Thackeray2022,Li2021}.
Our research aligns with these findings, identifying a relative intensity and probability increase of days with extreme rainfall under a future SSP3-7.0 scenario in Belgium at 15.4\% and 40.3\%, respectively. Note that future increases in the intensity and likelihood of extreme rainfall might even be larger than current model projections, since CMIP6 models may underestimate the historical scaling of precipitation extremes with global mean temperature~\citep{Kotz2024}.

The present study aimed to determine the dynamical and thermodynamical contributions to the changes in future extreme precipitation over Belgium.
Consistent with previous work (e.g.~\cite{Herrera-Lormendez2023, Otero2018, Kuttel2011, Tabarietal2020}), thermodynamics is found to be the primary attributor to future changes, driven by the intensification of the hydrological cycle under global warming~\citep{IPCC2023}. Rising temperatures enhance atmospheric water-holding capacity by approximately 7\% per °C, as described by the Clausius-Clapeyron relation~\citep{IPCC2023}, leading to increased evaporation and moisture availability, and consequently raising the potential for extreme precipitation events when the air mass cools and condenses.
This trend is particularly evident in mid-to-high latitudes, including Belgium, and results in a relatively uniform global fractional increase in extreme precipitation intensity across models when considered independently~\citep{Pfahl2017}. 
Dynamic factors, on the other hand, can offset the aforementioned effects in some areas (e.g.~Mediterranean) but do not appear for Belgium, aligning with e.g.~\cite{Tabari2020}. 

Our results demonstrate an increase in intensity and likelihood of extreme rainfall days under circulation types C, SW, and W (Fig.~\ref{fig_wtfrequency}).
The increase in likelihood is in line with for instance~\cite{Zappa2013}, who found an increase in the number of cyclones associated with extreme precipitation over the North Atlantic and Europe, while cyclonic occurrences were found to decrease. In line with our results (for W, NW, and SW), 
\cite{Herrera-Lormendez2023} found a projected likelihood decrease for the most important wet LWTs (C, W, NW, and SW) for summer over Europe, leading to the projected drying trend in Western-European summers following SSP5-8.5~\citep{Herrera-Lormendez2023}.

We used Eq.~\eqref{eq_decomposition} to decompose into a thermodynamic and a dynamic component and extended it to the case of extreme-precipitation intensity and probability. Our ``thermodynamic'' changes are defined as those changes that are not due to changes in weather types and may therefore not fully align with those in other works, especially those that are more process based. This is, however, a commonly-used simplification. In fact, all the processes that are responsible for a change in precipitation within a specific LWT are allocated to the thermodynamic term, while several non-thermodynamic processes are likely to occur. 
During dry LWTs, mesoscale features must be present on extreme precipitation days, that facilitate the initiation of convective processes, or at least processes capable of sustaining high rainfall rates over a prolonged period~\citep{Withford2024}. These features include a conditionally unstable air parcel, adequate low-level moisture, and an uplifting mechanism to carry moist air parcels to their level of free convection (\cite{Doswell1996}), processes that might change in a changing climate. Moreover, within WT changes in the temperature lapse rate~\citep{OGorman2009} and vertical wind velocities~\citep{Pfahl2017} all relevant for precipitation, might occur. 
Some improvements would arise in future work from including the vertical structure of the LWTs ~\citep{Herrera-Lormendez2023}. 
As noted by~\cite{Nie2018}, increased latent heating from extreme rainfall can modify associated atmospheric circulations, affecting both the strength and vertical structure of large-scale vertical motion. Including 
vertical structures in WT, could reveali how latent heating influences circulation patterns in extreme events~\citep{Nie2018}.

The dynamical term takes into account only the changes in the probability of large-scale LWTs. A poleward expansion of the tropical Hadley circulation in fall and winter~\citep{Grise2020} is often mentioned to be responsible for changes in LWT.
As a response, a poleward shift of westerly-like weather types will take place~\citep{Herrera-Lormendez2023} and increasing extreme precipitation.
Additionally, a stronger north-south gradient over the North Atlantic leads to stronger westerlies~\citep{Kjellström2018}, which was found be strongest in EC-EARTH- and CNRM-CM5-driven models. This could be an indication of the strong CNRM-CM6-1 signals for the strong (thermo)dynamic signals for intensity changes in C, that we found here.

It also needs to be acknowledged that dynamic and thermodynamic processes interact across different scales, and linking extreme precipitation events to a single atmospheric process is an oversimplification~\citep{Prein2023}.
For instance, rainfall extremes are influenced by the temperatures at which they occur~\citep{OGorman2009}.
Furthermore, seasonal differences further complicate the picture, as convective processes dominate in summer, while synoptic systems and large-scale ocean-to-land moisture transport play a more significant role in winter~\citep{Brogli2019}.

Our metric for extreme-rainfall intensity concerns the average excess rainfall on days of extreme rainfall. Another commonly used metric for rainfall intensity changes considers quantile changes (e.g. P99.9) between the future and historical period. This quantile-based rainfall intensity change is $11.9\%$ in our study (see Table~S1) which is comparable to our intensity change of 15.4\%.
However, as opposed to our intensity metric and as shown in Sect.~S2, this quantile-based metric is a monotonically increasing function of extreme-rainfall likelihood changes $\Delta P_{ext}(x_t)$. In other words, positive (negative) changes in $\Delta P_{ext}(x_t)$ give rise to positive (negative) changes in the quantile-based intensity change. Thus, the sign of dynamical or thermodynamical contributions (per LWT, per season, etc.) for $\Delta P_{ext}(x_t)$ will be the same for the quantile-based intensity changes. Additionally, the decomposition Eq.~\eqref{eq_decomposition} is not well defined for quantile changes as the function $G(d)$ (necessarily a function of the day), cannot be identified.

\section{Conclusion}\label{sec13}
This work investigates the dynamical contributions underlying extreme daily precipitation events in Belgium, exploring the historical and projected future changes associated with Lamb weather types (LWTs) using station observations and CMIP6-simulations.

Historical data indicate that wet cyclonic (C) weather types in fall and winter and westerly (W), and south-westerly (SW) in spring and summer are the primary drivers of extreme precipitation. Even though days with anticyclonic conditions occur most often, they rarely coincide with extreme rainfall. The GCM evaluation underscores significant differences in their ability to simulate weather types during all days and days with extreme precipitation, stressing the importance of a targeted model selection. By applying the Perkins Skill Score (PSS), 19 out of 24 models were retained for future projections, thereby excluding a model that was previously found as the best model at representing the average circulation probability distribution over Belgium.

The end-of-the-century projections under the SSP3-7.0 scenario show a substantial increase in the intensity and likelihood of extreme-rainfall events. The dynamic contributions are generally small and insignificant with respect to the thermodynamic changes. The exceptions to this occur in spring and summer when the dynamical changes in extreme-rainfall probability are significant and negative yet small, and, when the intensity changes contribute large uncertainties.

The results underscore the critical role of thermodynamic processes in shaping future precipitation extremes. Therefore, this work provides an incentive for the use of methods such as the pseudo-global-warming approach~\citep{schar1996surrogate} to probe the uncertainties related to the impact of climate change on extreme rainfall. Additionally, this work again highlights the necessity for targeted adaptation strategies to mitigate the growing risks associated with human-induced hydrological changes.

Further paths of interest include the investigation of the changes in extreme rainfall using projection data from regional climate models (RCMs) or convection-permitting models~\citep{Serras2024} rather than GCMs, the relation with temperature changes and the changes in rainfall extremes, 
the use of impact-tailored weather-type clustering methods (e.g.~\cite{Serras2024}), and, an extension of the framework in the context of attribution~\citep{vautard2016attribution} or subdaily rainfall extremes~\citep{whitford2024atmospheric}. These, however, are considered beyond the scope of the current work.

\section*{Acknowledgements}
We acknowledge Copernicus for the ERA5 reanalysis data. We also acknowledge the World Climate Research Programme, which, through its Working Group on Coupled Modelling, coordinated and promoted CMIP6. We thank the climate
model groups for producing and making available their model output, the Earth System Grid Federation (ESGF) for archiving the data and providing access, and the multiple funding agencies that support CMIP6 and ESGF.

\section*{Conflict of interest}
The authors declare no conflicts of interest.

\section*{Data availability}
The ERA5 reanalysis dataset can be accessed through the Copernicus Climate Change Service (C3S) at ECMWF (\url{https://cds.climate.copernicus.eu}). The CMIP6 model data can be accessed through the ESGF (\url{esgf-metagrid.cloud.dkrz.de/search}). The observed rainfall data can be downloaded from the opendata portal of the Meteorological Institute of Belgium (\url{opendata.meteo.be}).

\section*{Funding statement}
This research was supported by the Belgian Science Policy (BELSPO) under Contract B2/223/P1/CORDEXbeII.

\section*{Authors’ contributions}
\textbf{Jozefien Schoofs}: Writing – original draft; writing – review and editing; software; data curation; formal analysis; validation; investigation; visualization.
\textbf{Kobe Vandelanotte}: Methodology; writing – review and editing; software; data curation; visualization; conceptualization; supervision.
\textbf{Hans Van de Vyver}: Methodology; writing – review and editing; conceptualization; supervision.
\textbf{Line Van Der Sichel}: Software; data curation; formal analysis; validation; investigation.
\textbf{Matthias Vandersteene}: Software; data curation; formal analysis; validation; investigation.
\textbf{Fien Serras}: Methodology; writing – review and editing; software; data curation; visualization.
\textbf{Nicole P. M. van Lipzig}: writing – review and editing.
\textbf{Bert Van Schaeybroeck}: Methodology; writing – review and editing; data curation; visualization; conceptualization; supervision; project administration.

\section*{Supporting information}
Within the Supporting information, one may find:\\
Section S1: General nalytical derivation of Eq.~(2), i.e. the decomposition into thermodynamic and dynamic contributions.\\
Section S2: This section addresses the mathematical relation between the extreme-rainfall changes and quantile-based intensity changes.\\
Figure S1: Barplots representing the historical weather-type probability.\\
Figure S2: The PSS associated with the climatological LWT probability ($P_{clim}$) against the PSS score associated with the LWT probability during extreme-rainfall days ($P_{extr}$).\\
Figure S3: Total absolute future change in intensity and probability of extreme precipitation events.\\
Figure S4: Absolute future change in intensity and probability of extreme precipitation events per weather type.\\
Figure S5: Barplots showing the contributing terms to Eq.~(2) per weather type.\\
Figure S6: Showing the PSS score associated with the LWT probability $P_{cond}$ and the one associated with the climatological LWT probability ($P_{clim}$) versus the PSS score associated with $P_{extr}$.\\
Table S1: Overview of the CMIP6 models used in this project.

%\end{multicols}

\bibliography{sn-bibliography}

\newpage

\section*{S1 Decomposition into thermodynamic and dynamic contributions}\label{secS2}

Assume $G(d)$ is a quantity on day $d$ for which one wants to investigate the average change under climate change. This change is rewritten here as an exact sum of a \textit{thermodynamic} and a \textit{dynamic} contribution.

\subsection*{S1.1 Choice of $G(d)$ in case of extreme rainfall}
In our work, two cases are considered:
\begin{itemize}
	\item \textit{Probability of extreme rainfall}: $G(d)=0$ when rainfall on that day is below a fixed threshold and $G(d)=1$ otherwise. Averages are taken over all days of the considered period.
	\item \textit{Intensity of extreme rainfall}: $G(d)$ is equal to the rainfall excess on day $d$, above the threshold for extreme rainfall. Averages are only over the days with extreme rainfall.
\end{itemize}

\subsection*{S1.2 Rewriting the average in terms of weather types}
The average of $G(d)$ over the considered set of $D$ days of a period is then:
\begin{align}
	\overline{G} &= \frac{1}{|D|}\sum_{d\in D} C(d)
\end{align}
%%%%%%%%%%%%%%%%%%%%%%%%%%%%%%%%%%%%%%%%%%%%%%%%%%
%%%%%%%%%%%%%%%%%%%%%%%%%%%%%%%%%%%%%%%%%%%%%%%%%%
Note that, as aforementioned, $D$ may differ depending on the quantity considered. Each day $d$ can be categorized into a set of different weather types $i$. Therefore one can rewrite this sum over all days as a sum over all weather types $i$ and all days $d$ that belong to that weather type ($d\in i$):
\begin{align}
	\overline{G} &=\frac{1}{|D|}\sum_i\left( \sum_{d\in D_i}C(d)\right).
\end{align}
%%%%%%%%%%%%%%%%%%%%%%%%%%%%%%%%%%%%%%%%%%%%%%%%%%
%%%%%%%%%%%%%%%%%%%%%%%%%%%%%%%%%%%%%%%%%%%%%%%%%%%%
This can now simply be rewritten as follows by dividing and multiplying by $|D_i|$ which is the amount of days that a weather type $i$ occurs:
\begin{subequations}
	\begin{align}
		\overline{G}
		&=\sum_i \frac{|D_i|}{|D|} \left(\frac{1}{|D_i|}\sum_{d\in D_i}C(d)\right)\\
		&=\sum_i^G P_i \overline{G}_i\label{decomp}
	\end{align}
\end{subequations}
%%%%%%%%%%%%%%%%%%%%%%%%%%%%%%%%%%%%%%%%%%%%%%%%%%
%%%%%%%%%%%%%%%%%%%%%%%%%%%%%%%%%%%%%%%%%%%%%%%%%%%%%
Here we defined $P_i$ as the probability of weather type $i$ and $\overline{G}_i$ as the average of $C$ over the days of weather type $i$:
\begin{subequations}
	\begin{align}
		P_i&=\frac{|D_i|}{|D|},\\
		\overline{G}_i&=\frac{1}{|D_i|}\sum_{d\in i}C(d).
	\end{align}
\end{subequations}
%%%%%%%%%%%%%%%%%%%%%%%%%%%%%%%%%%%%%%%%%%%%%%%%%%%%%%%%
\subsection*{S1.3 \label{sec_otero}Decomposition}
We are interested in the climate change of the average $G$ i.e. the difference between the future ($f$) and historical ($h$) period:
\begin{align}
	\Delta\overline{G} &= \overline{G}^f - \overline{G}^h.
\end{align}
It follows from Eq.~\eqref{decomp} that:
\begin{align}
	\Delta\overline{G} &=\sum_i^G \left(P_i^f \overline{G}_i^f - P_i^h \overline{G}_i^h\right).
\end{align}
Using simple arithmetic, one can then write:
\begin{subequations}
	\begin{align}
		\Delta\overline{G} &=\sum_i^G \left[\left(P_i^f - P_i^h\right) \overline{G}_i^f + P_i^h \left(\overline{G}_i^f - \overline{G}_i^h\right)\right]\\
		&=\sum_i^G \Bigl(\,\underbrace{ \overline{G}_i^f \Delta P_i}_\text{Dyn.} + \underbrace{P_i^h \Delta \overline{G}_i}_\text{Thermodyn.}\Bigr),
	\end{align}
\end{subequations}
where we defined:
\begin{subequations}
	\begin{align}
		\Delta P_i &= P_i^f - P_i^h,\\
		\Delta \overline{G}_i &= \overline{G}_i^f - \overline{G}_i^h.
	\end{align}
\end{subequations}

\newpage

\section*{S2 On the definition of intensity changes for extreme rainfall}
\subsection*{S2.1 Introduction}
Here the connection is addressed between the frequency changes $\Delta P_{ext}(x_t)$ above a certain high threshold $x_t$ and the intensity changes under climate change. There are two ways to define the intensity changes: 1) \textit{The Peaks-over-Threshold (POT) based intensity change} $\Delta I_{POT}(x_t)$: This is the intensity definition used in this work and it is based on the average rainfall above a fixed threshold $x_t$. 2) \textit{The quantile-based intensity change} $\Delta I_{quant}(p_t)$: This is based on the quantile changes for a fixed high probability $p_t$. The goal here is to show that $\Delta I_{quant}(p_t)$ is a monotonically increasing function of $\Delta P_{ext}(x_t)$.

\subsection*{S2.2 Frequency changes}
We denote the rainfall cumulative distribution as $F$. Given a fixed high threshold $x_t$ (e.g.~$x_t=20$mm/day), the exceedance probability is, by definition $1- F(x_t)$.  
The probability changes $\Delta P_{extr}$ between the future (f) and historical (h) period are therefore
\begin{align}
	\Delta P_{extr}=(1-F_f(x_t))-(1-F_h(x_t)).
\end{align}
Alternatively one may also consider the relative likelihood change:
\begin{align}
	\Delta P_{extr,rel}=\frac{\Delta P_{extr}}{1-F_h(x_t)}.\label{rel_change}
\end{align}

\subsection*{S2.3 The POT-based intensity change $\Delta I_{POT}(x_t)$}
The POT-based intensity change considers the change in average rainfall above a fixed high threshold $(x_t)$.  For a sufficiently high threshold, the excess $y\equiv x- x_t$ is approximated by the Generalized Pareto (GP) distribution, which is of the form:
\begin{align}
	F(y|x>x_t)&=\frac{F(y)-F(x_t)}{1-F(x_t)}\approx 1-\left(1+\frac{\xi y}{\sigma}\right)^{-1/\xi}.\label{pot_cdf}
\end{align}
The average rainfall intensity $I_{POT}(x_t)$ above the  threshold $x_t$ is simply the first moment of the GP distribution: 
\begin{align}
	I_{POT}(x_t)&=\frac{\sigma}{1-\xi},\qquad\text{ if } \xi<1.
\end{align}
Under climate change, the intensity change is therefore:
\begin{align}
	\Delta I_{POT}&=\frac{\sigma_f}{1-\xi_f}- \frac{\sigma_h}{1-\xi_h}.
\end{align}
As the shape parameter often remains constant (e.g.~Dierickx, 2024), $\Delta I_{POT}\sim \sigma_f - \sigma_h$.

\subsection*{S2.4 The quantile-based intensity change $\Delta I_{quant}(p_t)$}
The quantile-based intensity change in extreme rainfall (not used in the manuscript) considers the change between future and historical quantiles, associated with a fixed high probability $p_t$ (e.g. $p_t=99\%$), and is given by:
\begin{align}
	\Delta I_{quant}(p_t) = F^{-1}_f(p_t)- F^{-1}_h(p_t)\label{inverse}
\end{align}

\subsection*{S2.5 Relation between POT-based and quantile-based intensity changes}
One can rewrite Eq.~\eqref{pot_cdf} as follows:
\begin{subequations}
	\begin{align}
		F(y)&=F(x_t)+[1-F(x_t)]F(y|x>x_t)\\
		&=F(x_t)+[1-F(x_t)]\left[ 1-\left(1+\frac{\xi y}{\sigma}\right)^{-1/\xi}\right]
	\end{align}
\end{subequations}
This allows us to find back the inverse function of $F(y)$ which is used in the quantile-based intensity change (Eq.~\eqref{inverse}):
\begin{align}
	F^{-1}(p_t)&=x_t+\frac{\sigma}{\xi}\left[\left(1-\frac{p_t-F(x_t)}{1-F(x_t)}\right)^{-\xi}-1\right]
\end{align}
The rainfall threshold $x_t$ so far is arbitrary. Assuming an increase in future extreme-rainfall likelihood ($F_f(x_t)>p_t$), we chose $x_t$ such that $F_h(x_t)=p_t$ readily finds that:
\begin{align}
	\Delta I_{quant}=\frac{\sigma_f}{\xi_f}\left[\left(1+\Delta P_{extr,rel}\right)^{\xi_f}-1\right].
\end{align}
with the relative likelihood changes $P_{extr,rel}$ defined in Eq.~\eqref{rel_change}.
In case of a decrease in future extreme-rainfall likelihood ($F_f(x_t)<p_t$), on the other hand, we chose $x_t$ such that $F_f(x_t)=p_t$ and find:
\begin{align}
	\Delta I_{quant}=-\frac{\sigma_h}{\xi_h}\left[\left(1+\Delta P_{extr,rel}\right)^{-\xi_h}-1\right].
\end{align}
From these results, it is clear that $\Delta I_{quant}$ is a monotonically increasing function of the relative likelihood change $\Delta P_{extr,rel}$ and therefore also as a monotonically increasing function of the extreme-rainfall likelihood changes $\Delta P_{ext}(x_t)$.

%To first order in $\Delta F(x_t)=F_f(x_t)- F_h(x_t)$, this gives
%\begin{align}
%\Delta I_{quant}=
%\frac{\sigma_f\Delta F(x_t)}{1-F_h(x_t)}
%\end{align}

\newpage

\section*{S3 Supplemantary Tables and Figures}\label{secS1}

\setcounter{figure}{0}
\renewcommand{\figurename}{Fig.}
\renewcommand{\thefigure}{S\arabic{figure}}

\begin{figure}[h]
	\centering
	\includegraphics[width=0.7\textwidth]{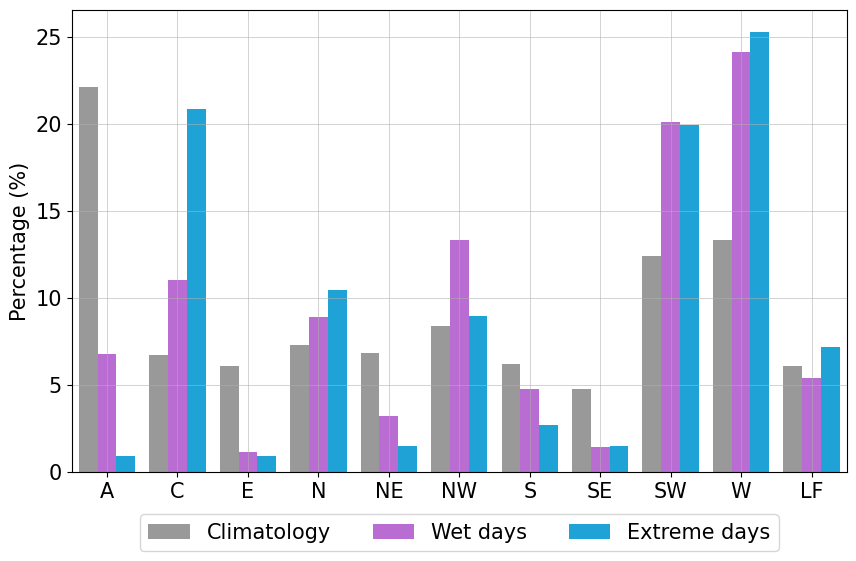}
	\caption{Barplots representing the weather-type probability of historical data using ERA5 MSLP data and station precipitation observation data (1951-2023). The grey bars indicate the probability of weather types for all days (i.e. the climatology) while the purple bars indicate the probability during days with precipiation (i.e. wet days), and the blue bars indicate the probability during days of extreme precipitation.}\label{fig_wetWT}
\end{figure}

\begin{figure}[h]
	\centering
	\includegraphics[width=0.7\linewidth]{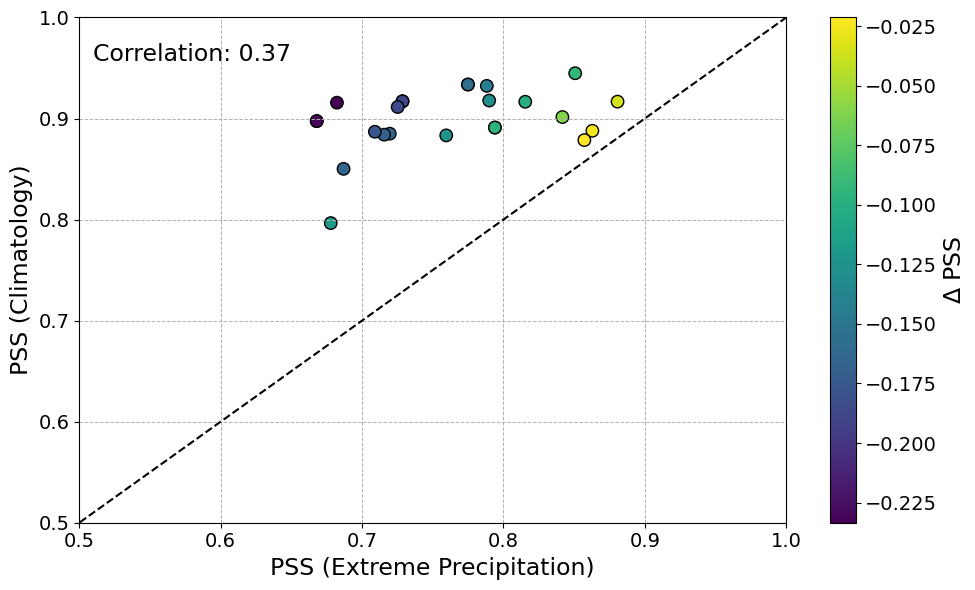}
	\caption{The PSS associated with the climatological LWT probability ($P_{clim}$) against the PSS score associated with the LWT probability during extreme-rainfall days ($P_{extr}$).}
	\label{fig_pss_extr_cf_clim}
\end{figure}

\begin{figure}[h]
	\centering
	\includegraphics[width=1\textwidth]{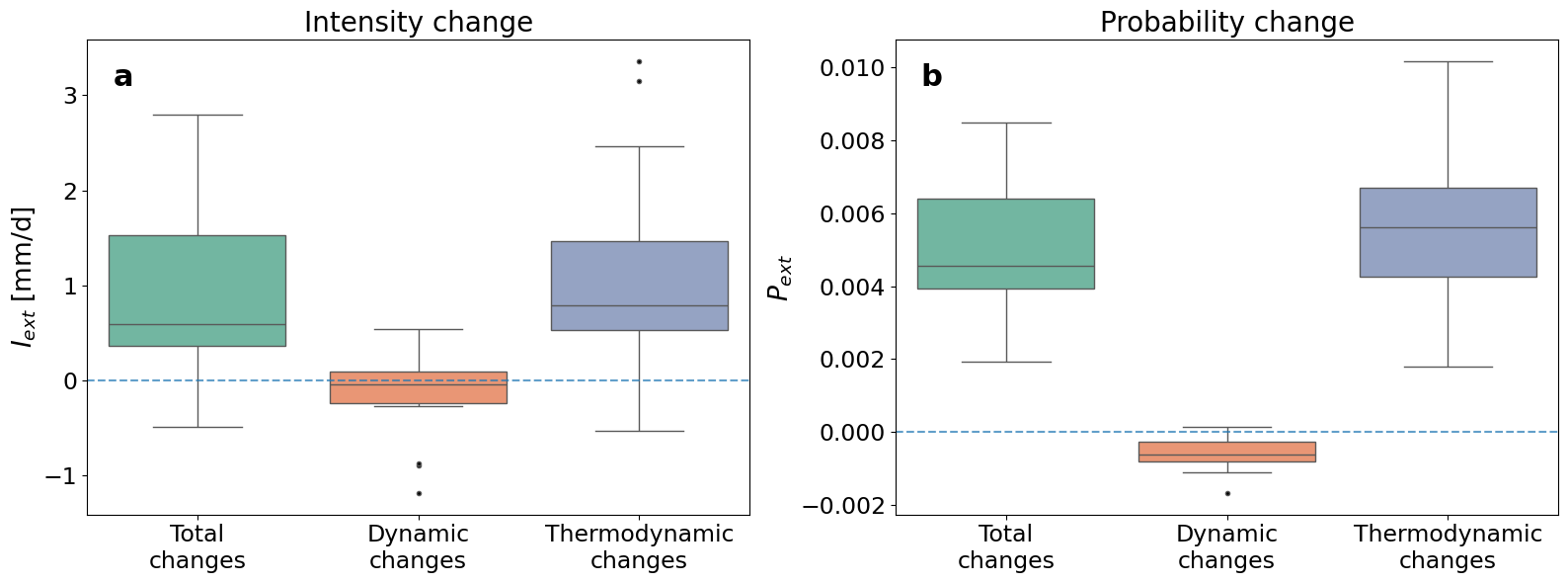}
	\caption{Change in intensity (left) and probability (right) for days with extreme precipitation for 2069-2098 with respect to 1985-2014, according to the selected 19 CMIP6 models. The green boxplot represents the total change in intensity, and the orange and blue indicate the dynamic and thermodynamic contributions, respectively. The median, interquartile range, and outliers of these changes over the CMIP6 models are represented by the solid black line, green box, and black dots, respectively.}\label{fig_decompositiontotal}
\end{figure}

\begin{figure}[h]
	\centering
	\includegraphics[width=1\textwidth]{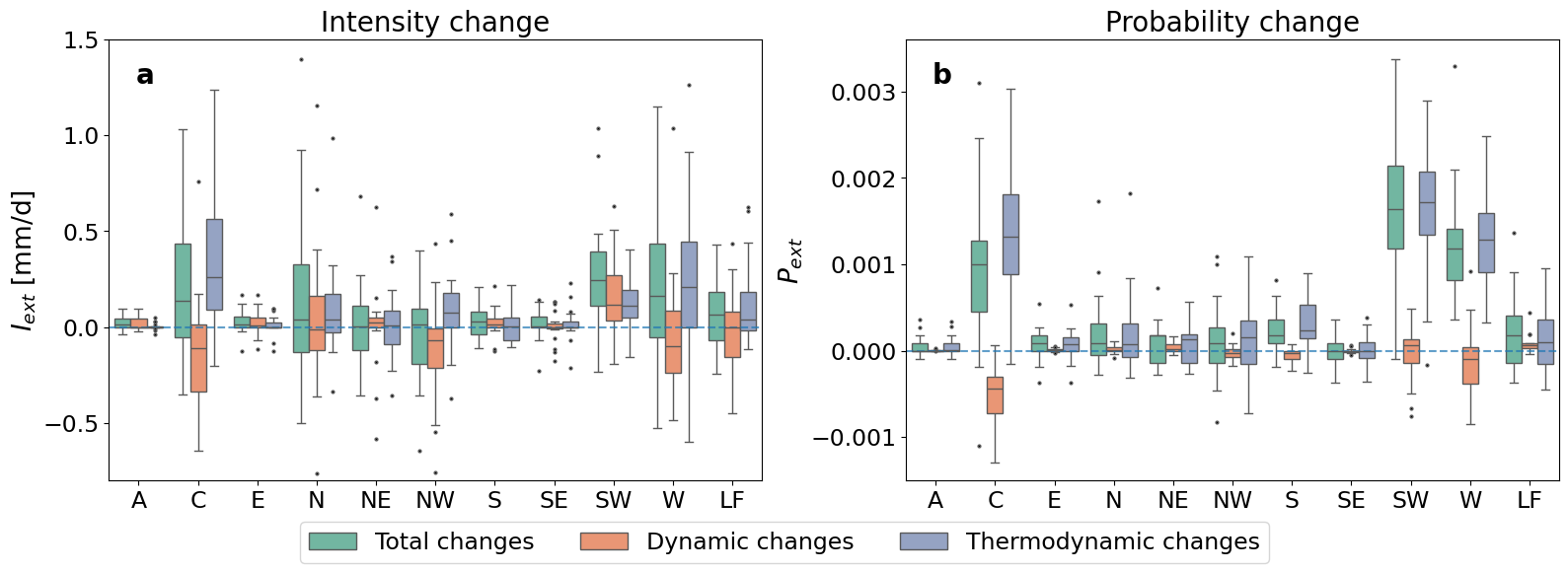}
	\caption{Change in intensity (left) and probability (right) of extreme precipitation events per weather type for 2069-2098 with respect to 1985-2014, averaged over 19 models within CMIP6. The green boxplot represents the total change in intensity, and the orange and blue indicate the dynamic and thermodynamic contributions, respectively. The median, interquartile range, and outliers of these changes over the CMIP6 models are represented by the solid black line, green box, and black dots, respectively. Note: the y-axis of the change in intensity is clipped between -0.8 and 1.5 to facilitate the differentiation between the different contributions, this caused 2 out-of-bound outliers for C (2.48mm/day for thermodynamic changes and -1.9mm/day for dynamic changes). The y-axis of the probability change is clipped between -0.0015mm/day and 0.0036mm/day, introducing out-of-bound outliers for W.}\label{fig_decompositionwt}
\end{figure}

\begin{landscape}
	\begin{figure}[ht]
		\centering
		\includegraphics[width=1.4\textwidth]{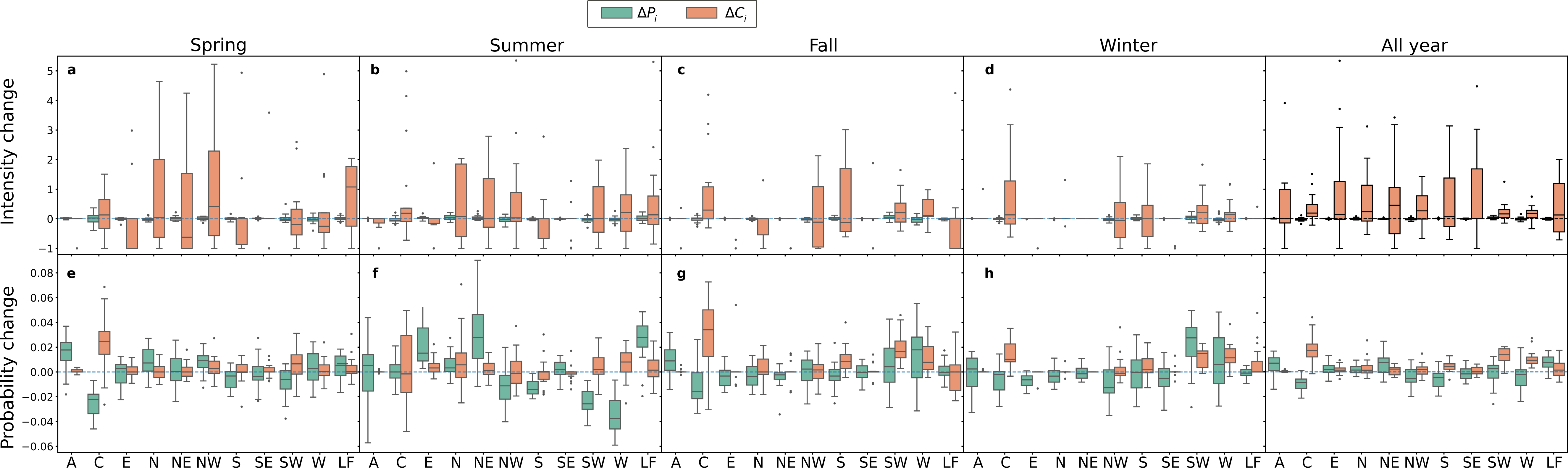}
		x\caption{Barplots showing the contributing terms to Eq.~(2).}\label{fig_modelevaluation}
	\end{figure}
\end{landscape}
\restoregeometry

\begin{figure}[h]
	\centering
	\includegraphics[width=0.7\linewidth]{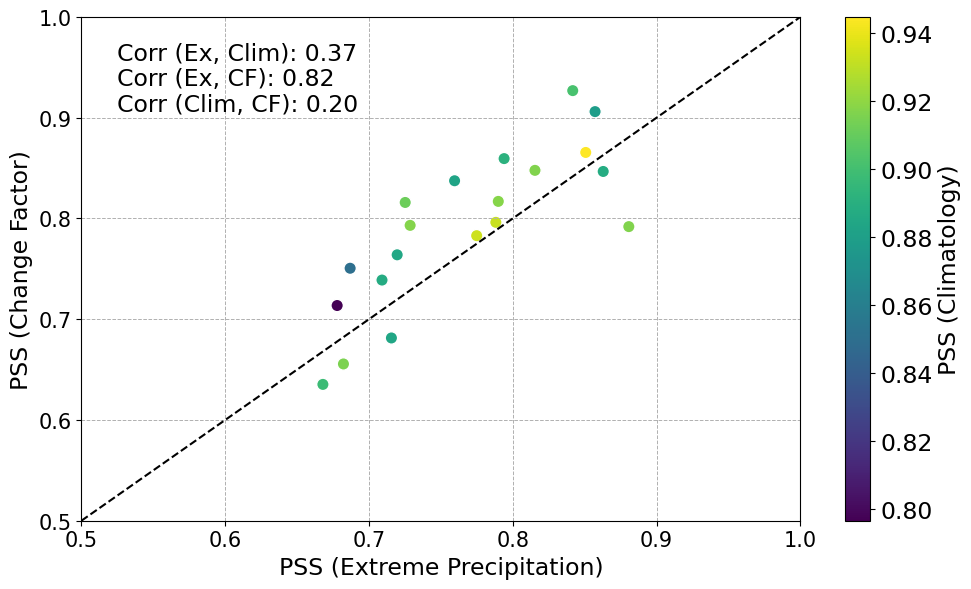}
	\caption{The PSS score associated with the LWT probability ($P_{cond}$, Change Factor or CF) versus the PSS score associated with $P_{extr}$.}
	\label{fig_pss_extr_clim}
\end{figure}

\setcounter{table}{0}
\renewcommand{\tablename}{Table}
\renewcommand{\thetable}{S\arabic{table}}

\newpage

\begin{table}[h]
	\caption{Overview of the CMIP6 models used in this project. The model name, run identifier, and institution/country are provided. The run column specifies the ensemble member used for each model, with the following run characteristics: initial conditions (r), initialization parameters (i), physics parameterizations (p), and forcings (f). The LWTs (per model) for which there are no days with extreme rainfall are given, for the historical (1985-2014) and future (SSP3-7.0; 2069-2098) periods. The last columns identify the historical and future value of the 98.87th percentile of daily precipitation (in mm). The 19 models retained in the analysis are written above the line. Below the line, the 5 low-performing models that are dropped are presented.}
	\label{tab:models_overview}%
	\begin{tabular}{@{}lllp{1.7cm}p{1.7cm}p{1.3cm}p{1.3cm}@{}}
		%\toprule
		\textbf{Model} & \textbf{Run} & \textbf{Institution/Country}  & \textbf{LWT missing (historical)} & \textbf{LWT missing}\newline \textbf{(future)} & \textbf{Historical threshold} & \textbf{Future threshold} \\
		%  & &    & Historical & Future \\
		%\midrule
		ACCESS-CM2           & r1i1p1f1 & CSIRO-ARCCSS/Australia & A & A & 20.11 & 22.26 \\
		ACCESS-ESM1-5        & r1i1p1f1 & CSIRO/Australia & SE & E & 21.67 & 24.66 \\
		CMCC-CM2-SR5         & r1i1p1f1 & CMCC/Italy & A, SE, S, E & A, SE & 19.04 & 21.18 \\
		CMCC-ESM2            & r1i1p1f1 & CMCC/Italy & SE, E & SE & 17.98 & 21.06 \\
		CNRM-CM6-1           & r1i1p1f2 & CNRM/France & LF, SE, E & & 15.54 & 17.04 \\
		CNRM-ESM2-1          & r1i1p1f2 & CNRM/France & A & NE & 16.59 & 17.11 \\
		EC-Earth3            & r1i1p1f1 & EC-Earth-Consortium &  & & 17.59 & 19.97 \\
		EC-Earth3-AerChem    & r1i1p1f1 & EC-Earth-Consortium & & A & 17.76 & 19.91 \\
		EC-Earth3-Veg        & r1i1p1f1 & EC-Earth-Consortium & A & & 16.84 & 19.93 \\
		EC-Earth3-Veg-LR     & r1i1p1f1 & EC-Earth-Consortium & & SE & 16.12 & 17.22 \\
		FGOALS-g3            & r1i1p1f1 & CAS/China & A, SE, NE & & 13.78 & 15.32 \\
		GFDL-ESM4            & r1i1p1f1 & NOAA-GFDL/USA & & & 17.26 & 18.28 \\
		IITM-ESM             & r1i1p1f1 & CCCR/India & A, NE & & 13.59 & 15.83  \\
		IPSL-CM6A-LR         & r1i1p1f1 & IPSL/France & E & SE & 22.53 & 26.10 \\
		MIROC6               & r1i1p1f1 & MIROC/Japan & A, SE, E & SE, E & 18.44 & 20.8 \\
		MPI-ESM1-2-HR        & r1i1p1f1 & MPI-M/Germany & A, SE, E & NE & 18.87 & 19.92 \\
		MPI-ESM1-2-LR        & r1i1p1f1 & MPI-M/Germany & A, SE, E & A, E & 18.66 & 20.72 \\
		MRI-ESM2-0           & r1i1p1f1 & MRI/Japan & SE & SE, E & 18.84 & 21.59 \\
		NorESM2-MM           & r1i1p1f1 & NCC/Norway &  & SE, NE, E & 15.94 & 17.33 \\
		\hline
		CanESM5              & r1i1p1f1 & CCCma/Canada & & & 16.22 & 19.34 \\
		TaiESM1              & r1i1p1f1 & AS-RCEC/Taiwan, China & & & 17.77 & 21.21 \\
		IPSL-CM5A2-INCA      & r1i1p1f1 & IPSL/France & & & 15.63 & 17.78 \\
		CESM2                & r4i1p1f1 & NCAR/USA & & & 17.04 & 17.83\\
		CESM2-WACCM          & r1i1p1f1 & NCAR/USA & & & 16.32 & 17.87
		%\bottomrule
	\end{tabular}
\end{table}

\end{document}